\documentclass[11pt,prd,showpacs,nofootinbib,oneside,hidelinks,a4paper,english]{revtex4-2}
\usepackage{graphicx}
\usepackage{amssymb}
\usepackage{amsmath}

\usepackage[english]{babel}
\usepackage{amsthm}
\usepackage{graphics}
\usepackage{dcolumn}
\usepackage{bm}
\usepackage{color}
\usepackage{epstopdf}
\usepackage{lipsum}
\usepackage{hyperref} 
\usepackage{float}
\usepackage{ulem}
\usepackage{tikz}
\usetikzlibrary{arrows.meta,positioning}
\usepackage[T1]{fontenc}
\usepackage[utf8]{inputenc}

\begin{document}
	\newcommand{\blue}[1]{\textcolor{blue}{#1}}
	\newcommand{\new}{\blue}
	
	\title{Low-Order Bessel-Type PID Dynamics in Lithium-Based Tritium Breeding and Heat-Removal Systems}
	
	\author{S. A. S. Borges}
	\author{S. D. Campos}\email{sergiodc@ufscar.br}
	\affiliation{Applied Mathematics Laboratory, DFQM/CCTS, Federal University of São Carlos, CEP 18052-780, Sorocaba, SP, Brazil}
	
	
	\begin{abstract}
		Lithium plays a dual role in deuterium–tritium fusion systems by enabling tritium breeding in blankets and providing an efficient heat-removal medium in liquid-metal components. However, most existing investigations treat neutronic behavior, jet thermohydraulics, and feedback control as largely decoupled layers, and there is currently no compact analytical framework that simultaneously links lithium-based tritium breeding, jet thermal response, and controller dynamics. In this work, we integrate nuclear cross-section data for deuterium–tritium and lithium reactions with a reduced thermohydraulic model of a liquid-lithium jet and an operator-theoretic formulation of feedback control. The resulting blanket/jet configuration should be interpreted as a conceptual, reduced-order demonstration of how two Li-based subsystems can be coupled in a unified analytical framework, rather than as a fully realistic reactor design in which an IFMIF-type neutron source is directly attached to a self-sufficient power blanket. We derive a low-order model describing jet thermal expansion under deuteron-beam loading and demonstrate that a continuous-time proportional–integral–derivative controller, expressed in operator form, can be locally embedded within a family of Bessel-type differential operators acting on the tritium-inventory tracking error. The results suggest that lithium-based breeding and heat-removal systems admit low-order, proportional–integral–derivative controllable dynamics that can be interpreted in terms of localized Bessel modes, providing a compact analytical framework for guiding future controller design and blanket/jet optimization.
	\end{abstract}

	\maketitle
	\section{Introduction}
	
	The climate crisis and the need to decarbonize the global energy mix motivate the search for energy sources with low greenhouse gas emissions and reduced production of long-lived hazardous waste \cite{iaea.2025,ciampichetti.2002}. Among the options under study, controlled nuclear fusion based on deuterium–tritium (D–T) reactions is widely seen as a promising path to large-scale, low-carbon power, offering high power density, limited long-lived radioactive waste, and a low risk of catastrophic accidents compared with conventional fission \cite{iaea.2025,ciampichetti.2002,nevins.1998,mitei.2024}. 
	
	In magnetically confined plasmas, such as those produced in tokamak devices, a toroidal vacuum vessel surrounded by electromagnetic coils that create a strong helical magnetic field to confine and heat the plasma \cite{hu.AAPPS.2023,ding.nature.2024,villari.2025,Solano.Plasm.Phys.2025}. In this configuration, the D-T reaction combines the largest fusion cross section at comparatively low ion temperatures with a favorable energy balance, making it the leading candidate for near-term fusion power plants \cite{nevins.1998,mitei.2024,bosch_hale_1992,Souza.arxiv.2019}. Achieving T self-sufficiency in these systems requires the integration of T-breeding blankets, in which lithium (Li) plays a dual role: it acts as a breeding material via neutron-induced reactions on $^6\mathrm{Li}$ and $^7\mathrm{Li}$, simultaneously serving as a medium for heat removal in liquid-metal blankets and high-speed Li jets \cite{sawan.2006,malang.2009,abdou.2021,lee.1972,manek.2023,morgan.2013,flament.1992,fukada.2010}. 
	
	Recent design advancements in large-scale magnetic confinement fusion programs underscore the increasingly stringent performance and reliability requirements placed on first-wall and blanket structural and functional materials. Notably, the revised International Thermonuclear Experimental Reactor (ITER) baseline configuration has replaced beryllium with tungsten as the first-wall material. This modification raises the kinetic energy threshold for D-induced surface damage and erosion, substantially suppresses physical sputtering, and correspondingly tightens the design and operational constraints on plasma-facing and blanket components with respect to plasma compatibility, thermal load management, and neutron-induced damage and activation control \cite{loarte.tech.report.2024,pitts.nucl.mat.2025}. Analogous design considerations are encountered in EU DEMO investigations, wherein multiple blanket concepts—in particular the Helium-Cooled Lithium Lead (HCLL), Helium-Cooled Pebble Bed (HCPB), and water-cooled lead–ceramic configurations (WCLL/WLCB)—systematically examine distinct combinations of liquid PbLi breeding materials, neutron multipliers, and advanced ceramic breeder systems \cite{HCLL_DEMO,HCPB_DEMO,WCLL_DEMO,morgan.2013_2,clark.2025}. In all these designs, Li-based components are central to closing the D-T fuel cycle and managing the high heat loads associated with fusion plasmas \cite{sawan.2006,malang.2009,abdou.2021,lee.1972,manek.2023,morgan.2013}.
	
	From a neutronics perspective, a principal performance metric for D–T fusion reactors is the T breeding ratio (TBR), defined as the ratio between the time-dependent T production rate in the breeding blanket and the T consumption rate as fusion fuel \cite{kuan.1999,zheng.2015,sawan.2006,malang.2009,abdou.2021,clark.2025}. While T self-sufficiency formally requires a minimum T breeding ratio of $\mathrm{TBR} \geq 1$, practical considerations, such as inventory losses, processing delays, and the need to maintain safety and operational margins—necessitate operating regimes in which the $\mathrm{TBR}$ exceeds unity \cite{kuan.1999,zheng.2015,sawan.2006,malang.2009,abdou.2021}. Consequently, many DEMO and ARIES design studies adopt a calculated TBR of approximately 1.10 as a pragmatic target, thereby providing a margin of roughly 10$\%$ above unity to compensate for tritium decay and processing losses, deficiencies and uncertainties in nuclear data and neutronics modeling, as well as port and penetration effects \cite{abdou.2021}. More generally, design targets in the range $\mathrm{TBR} \approx 1.05-1.15$ are often considered, corresponding to an excess of about 5–10$\%$ above unity, in order to accommodate T losses, processing delays, and uncertainties associated with nuclear data and three-dimensional blanket modeling \cite{abdou.2021,malang.2009,hadi.2026}.
	
	Comprehensive Monte Carlo neutronics and burnup analyzes demonstrate that TBR exhibits a nonlinear dependence on blanket geometry, material composition, and neutron energy spectra, and that securing sufficient breeding margins within realistic blanket configurations constitutes a stringent and nontrivial constraint on overall reactor feasibility \cite{kuan.1999,zheng.2015,sawan.2006,malang.2009,abdou.2021,morgan.2013_2,clark.2025,abdou.2021}.
	
	In parallel, Li also appears in high-intensity neutron-source concepts based on D beams incident on high-speed liquid Li jets, such as the International Fusion Materials Irradiation Facility - DEMO Oriented NEutron Source (IFMIF-DONES) \cite{knaster.2016,dovidio.2023,hassanein.1996,moslang.2008}. In these facilities, the D-Li stripping reaction is used to generate neutron spectra similar to those from D-T fusion, exposing test materials to fusion-relevant conditions \cite{knaster.2016,dovidio.2023,hassanein.1996,moslang.2008}. The free-surface Li jet in IFMIF-type systems must withstand substantial energy and momentum deposition from the D beam, which drives strong thermal gradients, surface perturbations, and flow instabilities that place additional design and operating constraints on the target and associated heat-removal systems \cite{knaster.2016,dovidio.2023,hassanein.1996,moslang.2008,flament.1992}.
	
	Because both tokamak blankets and Li-jet targets must operate near design points with small, controlled deviations in TBR, temperature, and flow parameters, feedback control strategies are essential to keep key observables within acceptable bounds \cite{morgan.2013_2,clark.2025,morgan.2013,Astrom_2002,morgan.2013}. Previous work on T breeding control within liquid-metal blankets has shown that relatively simple proportional–integral–derivative (PID) controllers can regulate the $^6$Li:$^7$Li enrichment ratio and maintain TBR near a target value, based on local linearizations of more complex neutronics models \cite{morgan.2013_2,clark.2025,morgan.2013,morgan.2013}. However, most of the existing literature treats neutronics, thermohydraulics, and feedback control as largely separate layers. There is currently no compact analytical framework that connects Li-based T breeding, jet thermohydraulics, and PID control in a way that illuminates the underlying dynamical structure and can guide controller tuning and system design \cite{morgan.2013_2,clark.2025,morgan.2013,Astrom_2002,morgan.2013}.
	
	In this work, we address this gap by combining nuclear data for D-T and Li reactions with a reduced-order thermohydraulic model of a liquid Li jet, complemented by an operator-theoretic formulation of the feedback control system. First, we review the basic physics of D-T fusion and the principal Li-based nuclear reactions responsible for T production, emphasizing their impact on TBR and blanket performance \cite{kuan.1999,zheng.2015,sawan.2006,malang.2009,abdou.2021,lee.1972,manek.2023,bosch_hale_1992,wielunska_2016,Souza.arxiv.2019,Wang.Comm.Phys.2019,abdou.2021}. Second, we formulate a reduced-order thermohydraulic model for jet thermal expansion based on coupled mass and heat transport equations that govern the response of a liquid Li jet to energy and momentum deposition from a D beam \cite{knaster.2016,dovidio.2023,hassanein.1996,moslang.2008,flament.1992,fukada.2010}. Third, we introduce the continuous-time PID controller in operator form and show that, after localization around a reference operating point, it is locally equivalent to a Bessel-type differential operator acting on the T-inventory error \cite{arfken_livro,Astrom_2002,morgan.2013}. By constructing a simple second-order error-dynamics model, we derive an explicit correspondence between the PID gains and effective Bessel parameters and illustrate this mapping with numerical examples \cite{arfken_livro,Astrom_2002,morgan.2013}.
	
	Our results indicate that Li-based breeding and heat-removal systems exhibit low-order. These PID-controllable dynamics can be interpreted in terms of localized Bessel modes, providing a unified perspective on T breeding, thermal management, and feedback control in Li-based fusion technologies \cite{knaster.2016,dovidio.2023,morgan.2013_2,clark.2025,morgan.2013,arfken_livro,Astrom_2002,morgan.2013}. It should be explicitly noted, however, that the configuration considered in this study is conceptual in nature and does not constitute a direct proposal for a realistic fusion power-plant architecture. Rather, we synthesize features inspired by DEMO-relevant T-breeding blankets and IFMIF-DONES-type liquid Li jets to define a simplified, hypothetical system. This reduced model is employed as an analytical testbed for investigating the coupled interactions among T-breeding, thermal management, and feedback control.
	
	The paper is organized as follows. Section \ref{sec:intro} reviews the basic physics of D-T fusion and the main Li-based T-breeding reactions. Section \ref{sec:jet} presents the reduced description of jet thermal expansion under D-beam loading (Appendix~A provides additional verification and benchmark results for this reduced thermohydraulic model). Section \ref{sec:pid} introduces the PID operator and the Bessel differential recurrences and develops the PID–Bessel correspondence. Section \ref{sec:tbr} connects this framework to the TBR. Section \ref{sec:jet_control} illustrates the ideas with a reduced jet/blanket–PID model, and Section \ref{sec:disc} summarizes our conclusions and discusses prospects for future work.

	\section{D-T fusion}\label{sec:intro}
	
\begin{figure}
		\centering
		\includegraphics[width=0.6\linewidth]{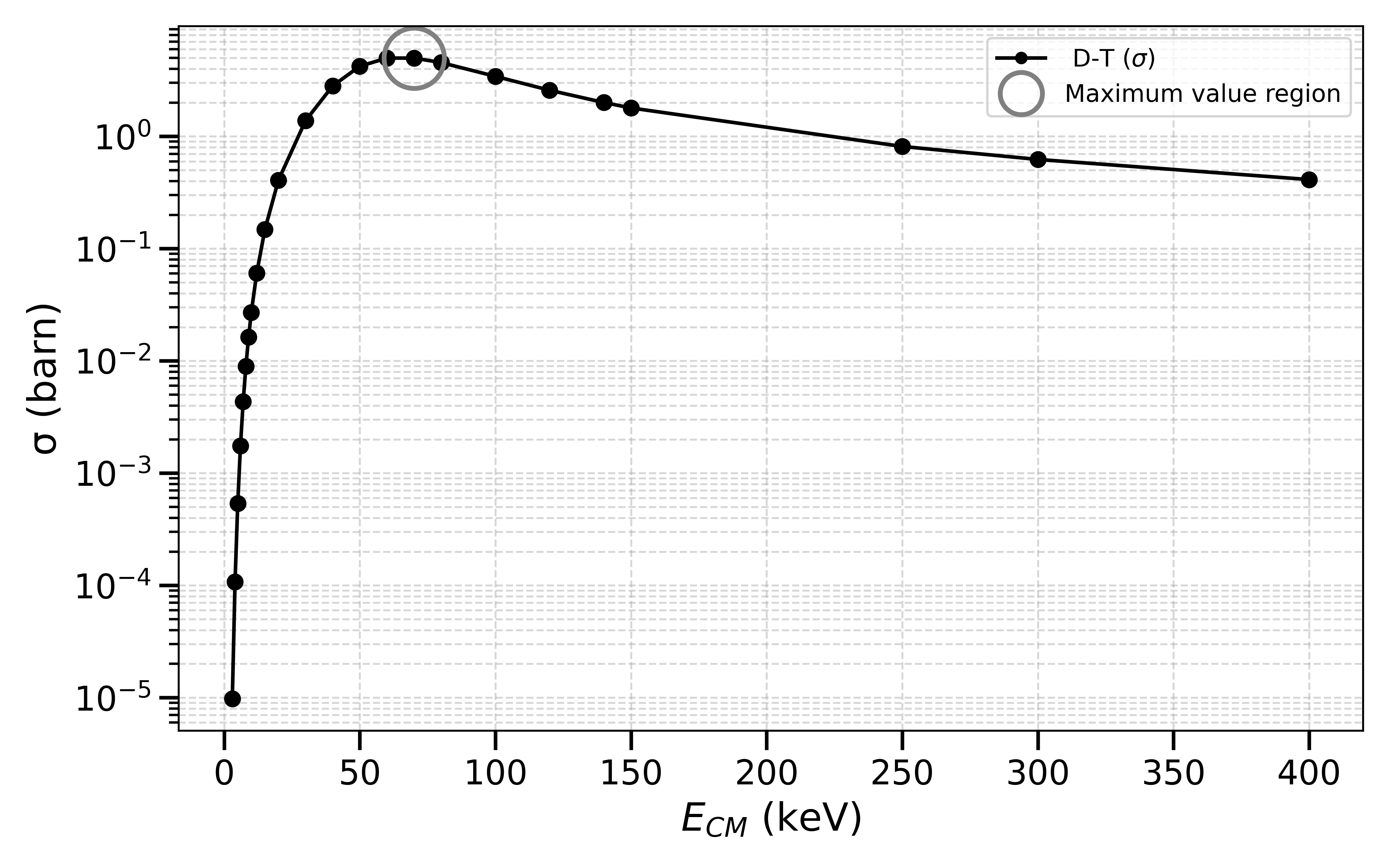}
		\caption{Total cross section for D–T collisions as a function of the center-of-mass energy $E_{CM}$. The dark gray circle marks the maximum $\sigma \approx 5.0$ barns at $E_{CM} \approx 65$ keV, a range compatible with ion temperatures envisaged for near-term D–T reactors. Experimental data are from Ref.~\cite{bosch_hale_1992}.}
		\label{fig:DT_sigma}
	\end{figure}

	The most relevant fusion reaction for energy applications is D-T fusion,
	\begin{equation}\label{eq:DT}
		{}^{2}\mathrm{H} + {}^{3}\mathrm{H} \rightarrow {}^{4}\mathrm{He} + n + 17.6~\mathrm{MeV},
	\end{equation}
	which combines a large cross-section at relatively low energies with a favorable energy balance, making it the natural candidate for the first commercial fusion reactors. Figure \ref{fig:DT_sigma} shows the total cross section, $\sigma$(barn), for several center-of-mass energies, $E_{CM}$ (keV), for D-T collisions. The maximum value of $\sigma_{max}$ ($\approx$ 5.0 barns) is achieved at relatively low energies ($\approx 65$ keV). The combination of the great value for the $\sigma$ at low energies is especially interesting when compared to the D-${}^3\mathrm{He}$, where $\sigma_{max}\approx 0.8$ barn at $E_{CM}\approx 600$ keV \cite{wielunska_2016}. 
	
	However, T is practically absent in nature because it is
	radioactive and has a relatively short half-life of 12.32 years \cite{abdou.2021}. Tritium can be generated by the interaction of neutrons and Li, produced \emph{in situ}, typically from reactions involving Li in breeding blankets through two main reactions 
	\begin{eqnarray}\label{eq:reactions}
		{}^6\mathrm{Li}+n \rightarrow {}^3\mathrm{H}+ {}^4\mathrm{He}+4.8\, \mathrm{MeV},\,\, {}^7\mathrm{Li}+n \rightarrow  {}^3\mathrm{H}+ {}^4\mathrm{He}+n-2.5\, \mathrm{MeV}
	\end{eqnarray}
	where the $Q$-values follow the usual convention that negative $Q$ corresponds to endothermic reactions. At the same time, 14.1 MeV neutrons carry most of the released energy and must be properly absorbed and converted into useful heat. In this context, Li plays a promising dual role: on the one hand, it acts as a T breeding material via reactions with neutrons \cite{morgan.2013}, and on the other hand, it can serve as a medium for heat containment and transport in the form of liquid metal in contact with the neutron flux and, in some advanced divertor concepts, directly with the edge plasma at plasma-facing components, where flowing liquid Li layers or films act as heat- and particle-flux removal media \cite{jiang.2025,sizyuk.2022}. Examples include liquid-Li divertor targets, Li-infused trench concepts, and open-surface Li flow divertors, in which a thin layer of liquid Li forms the plasma-facing surface and participates directly in the exhaust of heat and particles \cite{jiang.2025}. Moreover, concepts of intense neutron sources based on high-energy D beams incident on high-speed liquid Li jets have been proposed and studied as material test facilities for fusion reactors, exploiting the D–Li stripping reaction to generate neutrons with spectra similar to those from the D–T reaction \cite{hassanein.1996,moslang.2008}.
	
	One of the key performance parameters for D-T fusion reactors is the TBR. It is defined as the ratio between the time-dependent rate of T production within the breeding blanket, $\mathrm{TP}(t)$, of the reactor and the time-dependent rate at which T is consumed, $\mathrm{TC}(t)$, as fuel in the fusion reactions. Therefore, one writes
	\begin{eqnarray}
		\mathrm{TBR}(t)=\frac{\mathrm{TP}(t)}{\mathrm{TC}(t)}.
	\end{eqnarray}
	
	For self-sufficiency, the minimum required is $\mathrm{TBR}=1$. However, T losses are expected, and a more realistic estimate should account for uncertainties and/or deficiencies \cite{kuan.1999,zheng.2015,abdou.2021}. Then, a 10$\%$ safety margin can be used to account for uncertainties \cite{sawan.2006,malang.2009}. Considering reactions \eqref{eq:reactions}, TBR is nonlinear. It can be increased by optimizing the ${}^6\mathrm{Li}$ and ${}^7\mathrm{Li}$ atomic fractions for the neutron energy spectrum, and by increasing both the neutron flux and the breeding blanket volume. 
	
	\section{Jet Thermal Expansion}\label{sec:jet}
	
	The mathematical framework formulated in this work is not intended to represent any specific, currently anticipated fusion power plant concept in which an IFMIF-DONES-type D–Li neutron source is directly coupled to a T-breeding power blanket achieving net T self-sufficiency. Rather, it is constructed with reference to two well-established configuration classes that serve as motivating exemplars: (i) DEMO-relevant tokamak blanket systems, in which Li-based materials are employed both for T breeding and for heat removal, and (ii) liquid Li jet targets in IFMIF-DONES-type neutron sources, in which high-intensity D beams deposit power into a free-surface Li target. The reduced-order models developed here synthesize key features of these configurations to yield a compact analytical framework for Li-based T-breeding and thermal management systems operating under feedback control, rather than to propose a new, fully integrated reactor architecture.
	
	We here develop a reduced description of the thermal expansion of a liquid Li jet subjected to intense D-beam loading to obtain low-order observables that can later be used in feedback-control models \cite{knaster.2016,dovidio.2023}. The emphasis is not on reproducing the full three-dimensional computational fluid dynamics behavior of IFMIF-type targets, but rather on capturing the dominant thermohydraulic couplings in a form suitable for subsequent linearization and control design \cite{knaster.2016,dovidio.2023}.
	
	The IFMIF-DONES \cite{knaster.2016,dovidio.2023} target system employs a free‑surface liquid Li-jet about 25 mm thick and 260 mm wide, flowing at velocities of the order of 15 m/s in front of the D beam, with an average surface heat flux of several $10^8$ W/m$^2$ corresponding to a total deposited power of $\sim 5$ MW. In current EU DEMO studies, the main blanket concepts are: the HCLL blanket, using liquid PbLi as breeder and neutron multiplier; the HCPB blanket, using ceramic breeder pebble beds of lithiated compounds such as $\mathrm{Li_4SiO_4}$ and $\mathrm{Li_2TiO_3}$; and WCLL/WLCB variants, which combine Pb or PbLi multipliers with advanced ceramic breeders in tube or pebble bed configurations \cite{HCLL_DEMO,HCPB_DEMO,WCLL_DEMO}.
	
	In tokamak-based fusion power plant designs, Li is not employed as a high-energy beam incident on metallic targets, as is the case in IFMIF-type neutron source configurations. Instead, it is incorporated within the surrounding breeding blanket and, in certain designs, within liquid-metal or ceramic breeder components that interface with the plasma-facing structures. The fusion plasma itself consists predominantly of a D–T mixture, whereas Li is primarily utilized for T breeding and thermal energy extraction in the blanket region. Lithium is incorporated into the vessel structures, specifically within the breeding blanket that surrounds the plasma. Its principal functions in this configuration are (i) to capture the high-energy neutrons produced by the D-T fusion reactions to breed T {\it in situ}, and (ii) to participate in thermal management by facilitating the removal of heat, which can then be converted into electrical power.
	
	To model the coupling between the velocity field and the thermal expansion of the jet, we start from the mass-conservation equation, written as
	\begin{eqnarray}\label{eq:mass_cons}
		\frac{\partial \rho}{\partial t}+\nabla\cdot(\rho \vec{v})=0
	\end{eqnarray}
	where $\rho=\rho(x,y,z,t)$ is the mass density and $\vec{v}=(v_x,v_y,v_z)$ is the velocity distribution inside the flowing jet \cite{dovidio.2023,flament.1992,fukada.2010}. For the reduced model considered here, we assume that the jet is weakly compressible over the operating range of interest, so that density variations are small but not strictly zero and can be driven by temperature changes and beam-induced perturbations \cite{dovidio.2023,flament.1992,fukada.2010}. In addition, we adopt a simple parametrization for the transverse components of the velocity,
	\begin{eqnarray}\label{eq:vx_vy}
		v_x=k_xv, \qquad v_y=k_yv,
	\end{eqnarray}
	where $v$ is a characteristic jet-flow velocity and $k_x,\,k_y$ are dimensionless parameters that encode the relative magnitude of the transverse components. Substituting equation \eqref{eq:vx_vy} into equation \eqref{eq:mass_cons} and collecting terms, we obtain an equation for the longitudinal component $v_z$ of the velocity,
	\begin{equation}
		\frac{\partial \rho}{\partial t}
		+ k_x v_x \frac{\partial \rho}{\partial x}
		+ k_y v_y \frac{\partial \rho}{\partial y}
		+ \frac{\partial}{\partial z} \left( \rho v_z \right) = 0,
		\label{eq:continuity_vz}
	\end{equation}
	interpreted as governing the maximum perturbation velocity in the $z$-direction along the jet \cite{dovidio.2023,flament.1992,fukada.2010}.
	
	The evolution of the temperature field is described by the time-dependent heat-conduction equation with advection,
	\begin{equation}
		\rho C_p \left(
		\frac{\partial T}{\partial t}
		+ \vec{v} \cdot \nabla T
		\right)
		- \nabla \cdot \left( \lambda \nabla T \right)
		= q(x,y,z,t),
		\label{eq:heat}
	\end{equation}
	where $T = T(x,y,z,t)$ is the temperature, $C_p$ is the specific heat capacity, $\lambda$ is the thermal conductivity of Li, and $q$ is the volumetric heat source associated with energy deposition from the D beam \cite{dovidio.2023,flament.1992,fukada.2010}. In the IFMIF-DONES geometry, the jet can be approximated as a planar slab with thickness $h$ along one transverse direction and width $w$ in the orthogonal transverse direction, with the beam primarily depositing energy over a finite interaction length in $z$ \cite{knaster.2016,dovidio.2023}. For the reduced-order representation, equation \eqref{eq:heat} may be averaged over the thickness $h$ and/or the width $w$, thereby yielding an effective one-dimensional or quasi-two-dimensional formulation for $T(z,t)$ or $T(y,z,t)$, respectively. This procedure retains the leading-order interactions among beam loading, advective transport, and thermal conduction \cite{dovidio.2023,flament.1992,fukada.2010}. Here, “reduced-order” is used in the specific sense of analytical lumping and projection. Equations \eqref{eq:continuity_vz} and \eqref{eq:heat} are cross-sectionally averaged over the jet and then projected onto a low-dimensional set of physically motivated scalar observables $y_i(t)$, such as peak surface temperature and maximum longitudinal velocity perturbation. This yields a low-order linear time-invariant (LTI) model capturing the dominant thermohydraulic couplings in a form suitable for systematic linearization and control design. This meaning differs from data-driven surrogate modeling based on proper orthogonal decomposition, balanced truncation, or machine-learning regression on higher-fidelity datasets. Table \ref{tab:jet_params} presents the physical parameters used throughout this work.
	
	\begin{table}[t]
		\centering
		\caption{Representative input parameters for the reduced jet thermohydraulic model used in this work. The values are chosen to be consistent with IFMIF-DONES design studies and DEMO blanket concepts, and are intended to provide order-of-magnitude realism rather than to reproduce a specific design in detail.}
		\begin{tabular}{lccc}
			\hline\hline
			Parameter & Symbol & Value & Reference \\
			\hline
			Jet thickness & $h$ & $25~\mathrm{mm}$ & \cite{knaster.2016} \\
			Jet width & $w$ & $260~\mathrm{mm}$ & \cite{knaster.2016} \\
			Jet velocity & $v$ & $15~\mathrm{m/s}$ & \cite{knaster.2016} \\
			Surface heat flux & $q_{\text{surf}}$ & $3\times10^8~\mathrm{W/m^2}$ & \cite{hassanein.1996,moslang.2008} \\
			Total deposited power & $P_{\text{dep}}$ & $5~\mathrm{MW}$ & \cite{knaster.2016} \\
			Lithium density & $\rho$ & $0.51~\mathrm{g/cm^3}$ & \cite{flament.1992} \\
			Specific heat capacity & $C_p$ & $4.2\times10^3~\mathrm{J/(kg\,K)}$ & \cite{flament.1992} \\
			Thermal conductivity & $\lambda$ & $50~\mathrm{W/(m\,K)}$ & \cite{flament.1992} \\
			Interaction length & $L_{\text{int}}$ & $2$--$3~\mathrm{cm}$ & \cite{hassanein.1996} \\
			\hline\hline
		\end{tabular}
		\label{tab:jet_params}
	\end{table}
	
\begin{figure}
		\centering
		\includegraphics[width=0.8\linewidth]{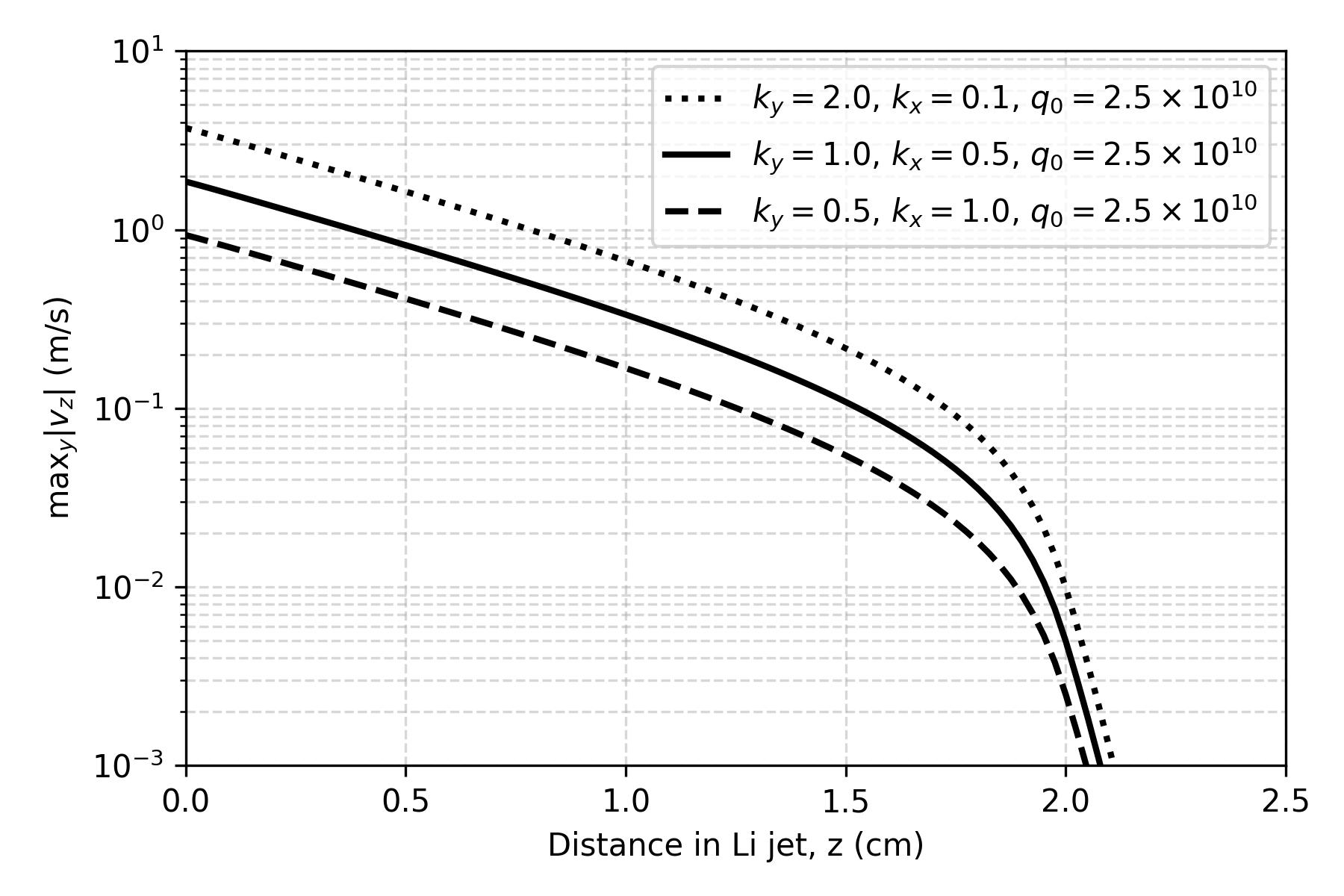}
		\caption{Different values for the maximum velocity $v_z$ (in logarithmic scale) toward the front surface.}
		\label{fig:plot5_6}
	\end{figure}

	For the illustrative computations shown in Figure \ref{fig:plot5_6}, we discretize the coupled mass-conservation and heat-conduction equations, \eqref{eq:continuity_vz} and \eqref{eq:heat}, on a one-dimensional grid along the longitudinal coordinate $z$, representing the beam–matter interaction region of the IFMIF-DONES liquid-metal jet. Spatial derivatives are approximated by second-order central finite differences on a uniform mesh, and the resulting semi-discrete system is advanced in time using an explicit first-order scheme with an adaptive time step selected to satisfy the Courant–Friedrichs–Lewy stability condition. At the jet inlet, we prescribe inflow velocity and temperature profiles consistent with the nominal IFMIF-DONES operating point, whereas at the outlet we impose advective (outflow) boundary conditions for both velocity and temperature. At the free surface exposed to the D beam, we enforce a mixed (Robin-type) thermal boundary condition that accounts for the balance between the imposed surface heat flux and convective heat removal. The volumetric heat source $q(x,y,z,t)$ is modeled as a Gaussian distribution in $z$ with a characteristic interaction length calibrated against IFMIF-DONES design parameters~\cite{knaster.2016,dovidio.2023}. Within this framework, the resulting reduced-order model furnishes a computationally tractable yet physically grounded description of the jet thermal response under beam loading.
	
	In particular, this figure presents the maximum longitudinal perturbation velocity $v_z$ (in logarithmic scale) toward the front surface as a function of the beam and flow parameters, thereby demonstrating that intense energy deposition gives rise to localized perturbations of velocity and temperature that remain confined within the beam–matter interaction region \cite{knaster.2016,dovidio.2023,flament.1992}. These perturbations introduce stringent constraints on the admissible operating regimes, including, for example, upper bounds on allowable surface temperature increases and on the maximum velocity deviation with respect to the mean flow \cite{knaster.2016,dovidio.2023,flament.1992}.
	
	The foregoing analysis demonstrates that the full fields $\rho(x,y,z,t)$ and $T(x,y,z,t)$, governed by equations \eqref{eq:continuity_vz} and \eqref{eq:heat}, can be systematically reduced to a low-dimensional set of scalar observables that are more directly pertinent to feedback-control design \cite{knaster.2016,dovidio.2023,morgan.2013_2,flament.1992}. Examples include the maximum jet temperature at the front surface, \(T_{\max}(t)\); the peak transverse or longitudinal velocity perturbation, \(v_{z,\max}(t)\); and integrated figures of merit that jointly characterize T-production efficiency and heat-removal performance \cite{knaster.2016,dovidio.2023,morgan.2013_2,clark.2025,morgan.2013}. Upon linearization of equations \eqref{eq:continuity_vz} and \eqref{eq:heat} about a steady operating point of the jet/blanket system and subsequent projection onto these observables, the resulting dynamics can be represented with high fidelity by low-order, LTI models over the relevant operating regime \cite{morgan.2013_2,clark.2025,morgan.2013}.
	
	The numerical implementation of equation \eqref{eq:heat} was verified against analytical solutions for one-dimensional pure diffusion and against grid-refinement studies for a convection–diffusion test problem, confirming second-order spatial convergence and first-order temporal convergence, consistent with the chosen discretization scheme. In addition, for steady-state configurations with uniform volumetric heating, the numerical solution reproduces the expected linear (diffusion-dominated) or exponential (advection-dominated) temperature profiles within a few percent. Although the present model is too simplified to provide quantitative predictions for a specific IFMIF-DONES target, we verified that the predicted ranges of peak surface temperature rise and longitudinal velocity perturbations are consistent in order of magnitude with those reported in IFMIF-DONES thermohydraulic analyses~\cite{flament.1992,knaster.2016,dovidio.2023}. We emphasize that a full quantitative validation would require three-dimensional computational fluid dynamics with detailed nozzle and duct geometry, which lies beyond the scope of this work. Schematically, one can summarize this reduction chain as
	\begin{equation}
		\nonumber \bigl( \rho(x,y,z,t),\, T(x,y,z,t) \bigr)
		\;\rightarrow\;
		\{ y_i(t) \}
		\;\rightarrow\;
		\text{low-order LTI model}
		\;\rightarrow\;
		\text{PID/Bessel operator design},
		\label{eq:reduction_chain}
	\end{equation}
	where $\{ y_i(t) \}$ denotes a small set of scalar observables extracted from the full fields \cite{morgan.2013_2,clark.2025,morgan.2013,morgan.2013}. In the next section, we demonstrate that the standard PID controller, expressed in operator form, together with the differential recurrence relations for Bessel functions, furnishes a convenient abstract framework for characterizing the feedback shaping of low-order error dynamics. Within this framework, PID-based T and thermal control schemes can be understood as specific instances of a more general Bessel-type operator structure.
	
	The numerical implementation of the reduced thermohydraulic model has been verified against both a canonical pure-diffusion test and a convection–diffusion benchmark. For the diffusion-dominated configuration with uniform volumetric heating and Dirichlet boundary conditions, the discrete operator based on second-order central differences reproduces the analytical steady-state solution with relative errors down to the level of machine precision, as summarized in Appendix~A, Table~\ref{tab:diffusion_errors}. In addition, a convection–diffusion benchmark constructed to emulate the IFMIF-DONES beam–matter interaction zone has been analyzed via systematic grid-refinement studies, with the resulting error levels and scaling behavior reported in Appendix~A and Table~III. Together, these tests provide quantitative verification that the reduced jet model captures the leading-order thermohydraulic couplings in a numerically consistent manner and is suitable as a building block for the coupled jet/blanket–PID–Bessel framework.
	
	\section{PID operator and Bessel differential recurrences}\label{sec:pid}
	
	We recast the standard continuous-time PID law here in operator form and compare it with the differential recurrences of Bessel functions, thereby identifying a local correspondence between PID-based error dynamics and a family of Bessel-type operators.
	
	The continuous-time PID controller can be written as \cite{Astrom_2002,morgan.2013}
	\begin{equation}\label{eq:pid}
		u(t) = K\left(e(t) + \frac{1}{T_i}\int_0^t e(\tau)\,d\tau
		+ T_d\frac{de}{dt}(t)\right),
	\end{equation}
	where $e(t)$ is the error signal, $K$ is the proportional gain, and
	$T_i$ and $T_d$ are the integral and derivative time constants, respectively. The mathematical structure of the PID equation \eqref{eq:pid} can be viewed as the action of a linear operator $C$ on $e$
	\begin{equation}
		u(t) = C[e](t),
	\end{equation}
	with
	\begin{equation}\label{eq:c_def}
		C :=  K\left(I + \frac{1}{T_i}\,\mathcal{I} + T_d\,\mathcal{D}\right),
	\end{equation}
	where
	\begin{equation}
		\nonumber I[e](t) = e(t),\qquad
		\mathcal{I}[e](t) = \int_0^t e(\tau)\,d\tau,\qquad
		\mathcal{D}[e](t) = \frac{de}{dt}(t).
	\end{equation}
	
	On the other hand, consider the Bessel functions of the first kind $J_\nu(x)$, which satisfy the standard differential-recurrence relations (for real $\nu$) \cite{arfken_livro}
	\begin{equation}\label{eq:bessel-rec1}
		J_{\nu-1}(x) + J_{\nu+1}(x) = \frac{2\nu}{x}J_\nu(x),
	\end{equation}
	\begin{equation}\label{eq:bessel-rec2}
		2\,J'_\nu(x) = J_{\nu-1}(x) - J_{\nu+1}(x),
	\end{equation}
	where $'$ denotes differentiation with respect to $x$.
	Equations \eqref{eq:bessel-rec1} and \eqref{eq:bessel-rec2} can be written in operator form by introducing the differential operator
	\begin{equation}\label{eq:def_1}
		\mathcal{L}_\nu
		:= \frac{d}{dx} - \frac{\nu}{x},
	\end{equation}
	acting on the family $\{J_\nu\}_{\nu\in\mathbb{R}}$. Using equations
	\eqref{eq:bessel-rec1} and \eqref{eq:bessel-rec2}, and definition \eqref{eq:def_1}, one writes, respectively,
	\begin{eqnarray}
		\nonumber J_{\nu+1}(x) &=& \left(\frac{\nu}{x} - \frac{d}{dx}\right)J_\nu(x) = -\mathcal{L}_\nu[J_\nu](x),\\
		\nonumber J_{\nu-1}(x) &=& \left(\frac{d}{dx} + \frac{\nu}{x}\right)J_\nu(x)=\mathcal{L}_\nu[J_\nu](x).
	\end{eqnarray}
	
	Analogously to the manner in which a PID controller generates a new control signal $u$ from the error signal $e$ via a fixed linear combination of derivative, integral, and identity operators, the Bessel function recurrence relations generate neighboring functions $J_{\nu \pm 1}$ from $J_{\nu}$ through linear combinations of differentiation and multiplication operators.
	
	\subsection{Analogy Between the Two Structures}
	
	Consider $\mathcal{X}$ as a suitable function space in time (e.g., $L^2$ on $\mathbb{R}_+$) and $\mathcal{Y}$ as a function space in the spatial variable $x$. Moreover, consider that the PID operator $C:\mathcal{X}\to\mathcal{X}$ is defined by $C$ given by \eqref{eq:c_def} and the Bessel ``shift'' operator $S_\nu:\mathcal{Y}\to\mathcal{Y}$ can be defined by
	\begin{equation}
		S_\nu := -\mathcal{L}_\nu 
	\end{equation}
	acting on the Bessel family via
	\begin{equation}
		S_\nu[J_\nu](x) = J_{\nu+1}(x).
	\end{equation}
	
	Both $C$ and $S_\nu$ are linear operators built by combining primitive
	operators (differentiation and either integration or multiplication)
	with scalar coefficients. In particular, $C$ combines $I$, $\mathcal{I}$, and $\mathcal{D}$ with constants $K$, $K/T_i$, and $K T_d$, shaping the closed-loop dynamics through proportional, integral, and derivative action. In contrast, $S_\nu$ combines $d/dx$ and multiplication by $\nu/x$ to ``shift'' the order of the Bessel function while preserving the underlying Bessel differential equation structure.
	
	Formally, one can see both as elements of suitable operator algebras:
	\begin{equation}
		C \in \operatorname{Alg}\{I,\mathcal{I},\mathcal{D}\},\qquad
		S_\nu \in \operatorname{Alg}\Bigl\{\frac{d}{dx},\,M_{1/x}\Bigr\},
	\end{equation}
	where $M_{1/x}$ denotes multiplication by $1/x$. The structural analogy
	is that both control action (via $C$) and order-shifting in Bessel
	systems (via $S_\nu$) are realized through linear combinations of basic
	operators that act on a function space and generate families of related
	signals or modes.
	
	One writes the PID-based increment in the controller variable $r_\infty$ as \cite{morgan.2013}
	\begin{equation}\label{eq:PID-Morgan}
		\Delta r_\infty(t)\;=\;a\,E(t) + b\,\frac{dE}{dt}(t) + c \int_0^t E(\tau)\,d\tau,
	\end{equation}
	where $E(t)$ denotes the error (the difference between the target and measured T inventory), and $(a,b,c)$ are the proportional, derivative, and integral gains, respectively. Now, one defines the corresponding differential operator as
	\begin{equation}
		L_{\text{PID}} \;:=\;b\,\frac{d}{dt} + a + c\,\mathcal{I},
	\end{equation}
	where $\mathcal{I}[E](t) := \int_0^t E(\tau)\,d\tau$ is the causal integration operator acting on $E$. Consider now the classical Bessel operator \cite{arfken_livro}
	\begin{equation}\label{eq:bessel_op}
		L_\nu \;:=\; x^2 \frac{d^2}{dx^2} + x \frac{d}{dx} + (x^2 - \nu^2),
	\end{equation}
	acting on sufficiently smooth functions of $x>0$, where $\nu\in\mathbb{R}$ is a fixed order parameter. Let $x_0>0$ be a reference point and assume that $x$ is confined to a small neighborhood of $x_0$. Then, in this neighborhood, $L_\nu$ admits the local approximation
	\begin{equation}\label{eq:local-Bessel-Morgan}
		L_\nu \;\approx\;x_0^2 \frac{d^2}{dx^2} + x_0 \frac{d}{dx}
		+ (x_0^2 - \nu^2),
	\end{equation}
	which is a second-order linear differential operator with constant
	coefficients. Furthermore, one writes $x = \alpha t$ with $\alpha>0$ and defines
	$y(x) := E(t)$ with $t = x/\alpha$. Under this change of variables, equation \eqref{eq:local-Bessel-Morgan} becomes, up to an overall scaling factor,
	\begin{equation}\label{eq:model_order}
		\tilde L_\nu
		\;\propto\;
		\tilde a_2 \frac{d^2}{dt^2}
		+ \tilde a_1 \frac{d}{dt}
		+ \tilde a_0,
		\qquad
		\tilde a_2 \propto x_0^2,\;
		\tilde a_1 \propto x_0,\;
		\tilde a_0 \propto x_0^2 - \nu^2.
	\end{equation}
	
	One can choose $(a,b,c)$ and the scaling $\alpha$ such that the first-order representation of the PID operator \eqref{eq:PID-Morgan} (viewed at the level of its underlying second-order error dynamics) coincides, up to scaling, with the localized Bessel-type operator $\tilde L_\nu$. In this sense, the PID increment can be embedded as a special case of a Bessel-type operator acting on the error trajectory $E(t)$. 
	
	It is important to emphasize that the foregoing identification is not intended to establish a strict global equivalence between the complete Bessel–Sturm–Liouville operator and the discrete PID controller \cite{Astrom_2002,morgan.2013}; their spectral formalisms remain distinct. Instead, it demonstrates that after localization around a reference point $x_0$ (freezing the coefficients at $x_0$) and a suitable rescaling $x=\alpha t$, the effective operator governing the PID increment \eqref{eq:PID-Morgan} can be viewed as an element of a Bessel-type family, with the controller gains $(a,b,c)$ playing the role of effective ``order'' and ``scale'' parameters associated with $(\nu,x_0)$. This provides a formal way to interpret the Li-ratio controller as implementing localized Bessel-type dynamics on the T inventory error.
	
	\subsection{A Simple Second-Order Error Model}
	
	A simple second-order error model can be constructed to match the PID gains $(a,b,c)$ \cite{morgan.2013} to the parameters of a localized Bessel operator. The first step is to consider a linearized second-order model for the T inventory error $E(t)$ of the form
	\begin{equation}\label{eq:error-second-order}
		E''(t) + a_1\,E'(t) + a_0\,E(t) = 0,
	\end{equation}
	where $a_0, a_1 > 0$ are the effective sensitivities of the blanket and purification system, respectively, to variations in the control variable. Now, one considers that the PID increment in the controller variable
	$r_\infty$ is given by  \eqref{eq:PID-Morgan}. Assuming that the physical system responds approximately proportionally to $\Delta r_\infty$ on a short time scale, and that the integral term contributes mainly to a slow drift of the operating point, we can associate equation \eqref{eq:PID-Morgan}, an effective second-order error dynamics of the form \eqref{eq:error-second-order}, with ($b\neq 0$)
	\begin{equation}\label{eq:a1-a0-from-ABC}
		a_1 = \frac{a}{b},\qquad
		a_0 = \frac{c}{b}.
	\end{equation}
	
	Identifying $x$ with time $t$ (or, more generally, taking $x=\gamma t$
	and absorbing $\gamma$ into a rescaling of coefficients), the
	associated homogeneous equation for $E(t)$ is
	\begin{equation}\label{eq:Bessel-like-error}
		E''(t)+\frac{1}{x_0} E'(t)+\Bigl(1-\frac{\nu^2}{x_0^2}\Bigr)E(t)= 0.
	\end{equation}
	
	Comparing \eqref{eq:Bessel-like-error} with the generic form
	\eqref{eq:error-second-order}, and considering identification \eqref{eq:a1-a0-from-ABC}, we can match PID gains to Bessel parameters, yielding the explicit correspondence ($a\neq 0$, $b\neq 0$)
	\begin{equation}\label{eq:x0-nu-from-ABC}
		x_0 = \frac{b}{a},
		\qquad
		\nu^2 = \Bigl(1 - \frac{c}{b}\Bigr)\frac{b^2}{a^2}.
	\end{equation}
	
	If $1 - c/b \ge 0$, then $\nu$ is real; otherwise, $\nu$ becomes
	complex, and the corresponding Bessel-type mode is exponentially
	modulated. In this way, for any choice of PID gains $(a,b,c)$ satisfying the conditions above, the effective second-order error dynamics can be
	viewed as the localization of a Bessel-type mode of order $\nu$ around
	the effective ``radius'' $x_0 = b/a$. This provides an explicit parameter map $(a,b,c)\;\longmapsto\;(x_0,\nu)$, realizing the PID controller as a special (localized) case of a Bessel-type operator acting on the error trajectory $E(t)$. Notice that in the equations (8)–(9), we use the conventional PID notation $(K, T_i, T_d)$, while equation \eqref{eq:PID-Morgan} we rewrite the same control law in terms of an equivalent gain triplet $(a,b,c)$ acting directly on the error $E(t)$. Specifically, one may identify
	\begin{eqnarray}
		a = K,\qquad b = K T_d,\qquad c = \frac{K}{T_i},
	\end{eqnarray}
	so that the proportional, derivative, and integral contributions of the standard PID controller are encoded in the coefficients multiplying $E(t)$, its time derivative, and its time integral, respectively. This reparameterization is introduced to emphasize the mapping between PID gains and the effective Bessel parameters $(x_0,\nu)$ in equation ~(24), without changing the underlying control action.
	
	To illustrate the mapping \eqref{eq:x0-nu-from-ABC}, consider a simple, dimensionless example in which the PID gains $(a,b,c)$ are chosen to be of the same order as typical values used in Ref. \cite{morgan.2013} (i.e., $a$ and $b$ of comparable size, and $c$ somewhat smaller, reflecting a less aggressive integral action). Then, for illustration, one assumes
	\begin{equation}
		a = 0.5,\qquad
		b = 1.0,\qquad
		c = 0.2,
	\end{equation}
	and using the explicit relations \eqref{eq:x0-nu-from-ABC}, one obtains the corresponding localized Bessel parameters 
	\begin{equation}
		x_0 = \frac{1.0}{0.5} = 2.0 \,\,\mathrm{and}\,\, \nu = \sqrt{3.2} \approx 1.78885. 
	\end{equation}
	
	Equivalently, this specific PID tuning may be interpreted as the selection of an effective Bessel-type mode of order $\nu \simeq 1.8$, spatially localized in the vicinity of the radius $x_0 \simeq 2$. We reiterate that this identification is intrinsically local in nature: the PID operator is derived as a second‑order, constant‑coefficient approximation of a Bessel‑type Sturm–Liouville operator in the neighborhood of a chosen reference point, rather than being established through a global spectral equivalence.
	
	During the time interval $t \in [t_1, t_2]$, the Bessel mode of order $\nu$ exhibits error dynamics that are accurately approximated by a second‑order model parameterized by the coefficients $(a_0,a_1)$, which, in turn, correspond to a set of effective PID gains $(a,b,c)$. Within this regime, the resulting closed‑loop T-dynamics are, to leading order, indistinguishable from those generated by a conventional PID controller.
	
	\section{Tritium Breeding Ratio}\label{sec:tbr}
	
	Throughout this work, the TBR is employed as a surrogate performance indicator within a reduced-order model that approximates DEMO-relevant blanket behavior. While the numerical case studies focus on operating conditions with TBR values marginally exceeding unity, this should be interpreted as an idealized control problem for Li-based breeding systems, rather than as a fully realistic power-plant design in which an IFMIF-type neutron source is coupled to a DEMO blanket. In a practical materials-testing facility, operation would instead be conducted in an experimental regime that does not impose the constraint of net T self-sufficiency.
	
	The TBR is suitable for connecting the neutronic behavior of Li-based blankets with the PID-Bessel framework developed above. One assumes that 
	\begin{eqnarray}\label{eq:n6_n7}
		n_6(t)+n_7(t)=n_{Li,tot}
	\end{eqnarray}
	where $n_6(t)$ and $n_7(t)$ are, respectively, the number density of $^6\mathrm{Li}$ and ${}^7\mathrm{Li}$ in the breeder region at time $t$. Furthermore, $n_{Li,tot}$ is the total Li density (or inventory) at time $t$, assumed to be constant.  The fraction of total Li that is 
	${}^6\mathrm{Li}$ at time $t$ is given by the enrichment ratio of ${}^6\mathrm{Li}$, written as
	\begin{eqnarray}\label{eq:enri_ratio}
		R(t)=\frac{n_6(t)}{n_6(t)+n_7(t)}\in [0,1],
	\end{eqnarray}
	and the error $E(t)$ is assumed to be proportional to $\mathrm{TBR}(t)-\mathrm{TBR}_{target}$. 
	
	It is important to stress that in detailed MCNP/FISPACT calculations,  TBR is a non-linear function of density, geometry, and neutron spectrum \cite{morgan.2013,clark.2025}. However, based on the operating regime relevant for feedback control, the dependence of TBR and impurity levels on the chosen control parameters (such as $^6\mathrm{Li}:{}^7\mathrm{Li}$ enrichment) is empirically close to linear, which motivates the use of linear controllers and local first-order (Taylor) approximations around a design point \cite{morgan.2013_2,Astrom_2002}. Therefore, one writes 
	\begin{eqnarray}\label{eq:tbr}
		\mathrm{TBR}(t)\approx \alpha R(t)+\beta,
	\end{eqnarray}
	where $\alpha$ is the sensitivity of $\mathrm{TBR}$ to changes in the ${}^6\mathrm{Li}$ enrichment near the operating point, whereas $\beta$ is the offset of the linear approximation.
	
\begin{figure}
		\centering
		\includegraphics[width=8.0cm,height=5.0cm]{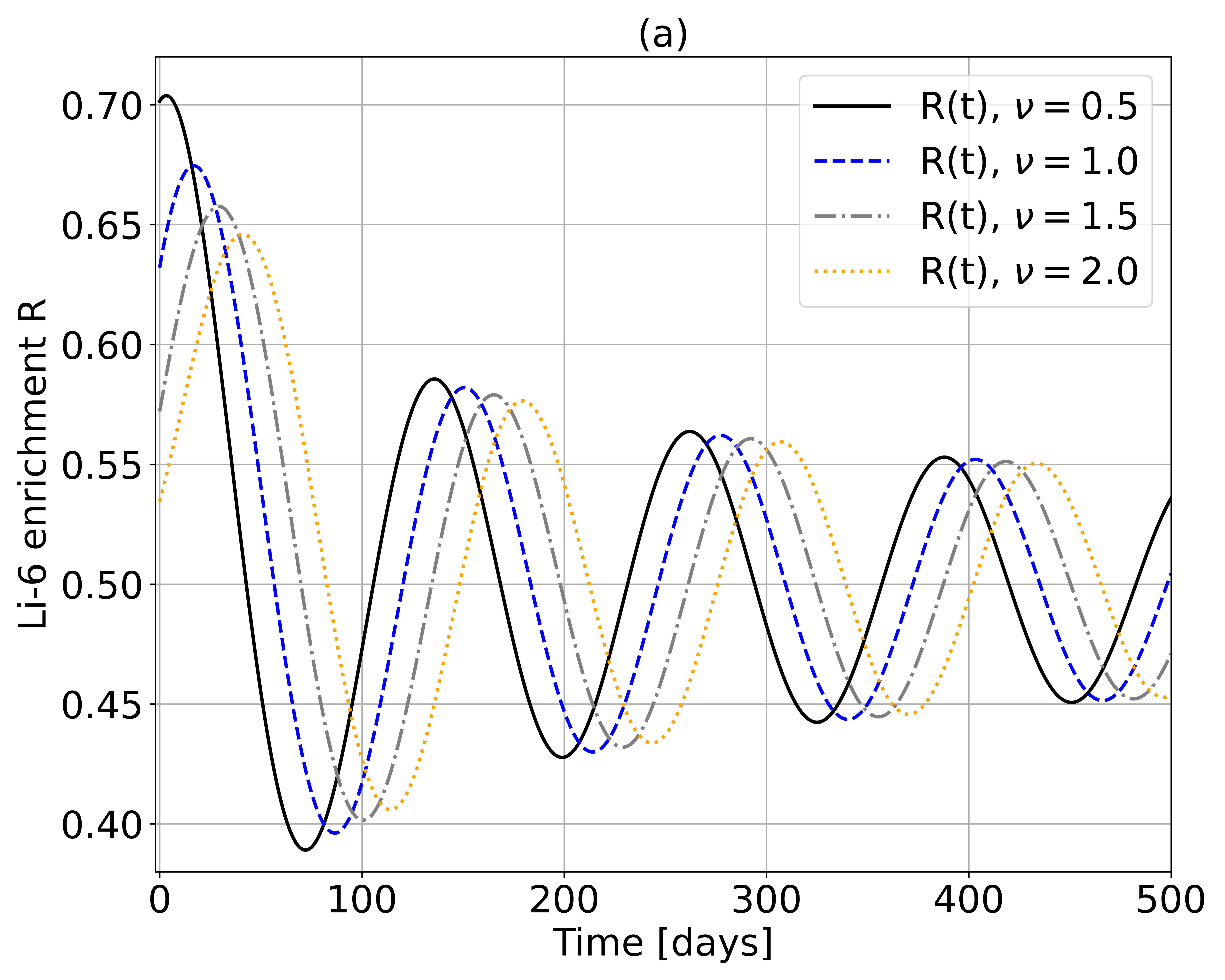}
		\includegraphics[width=8.0cm,height=5.0cm]{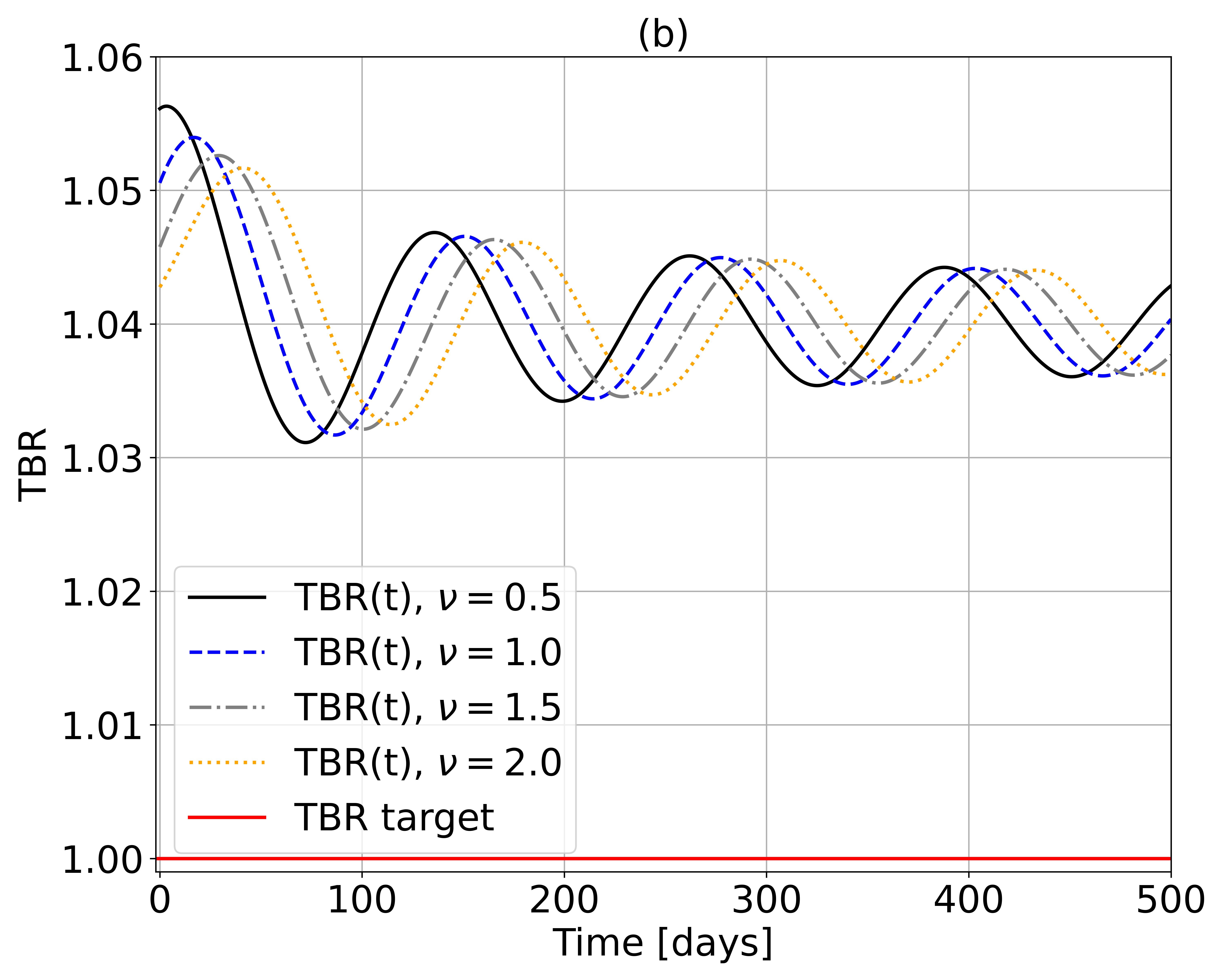}
		\includegraphics[width=8.0cm,height=5.0cm]{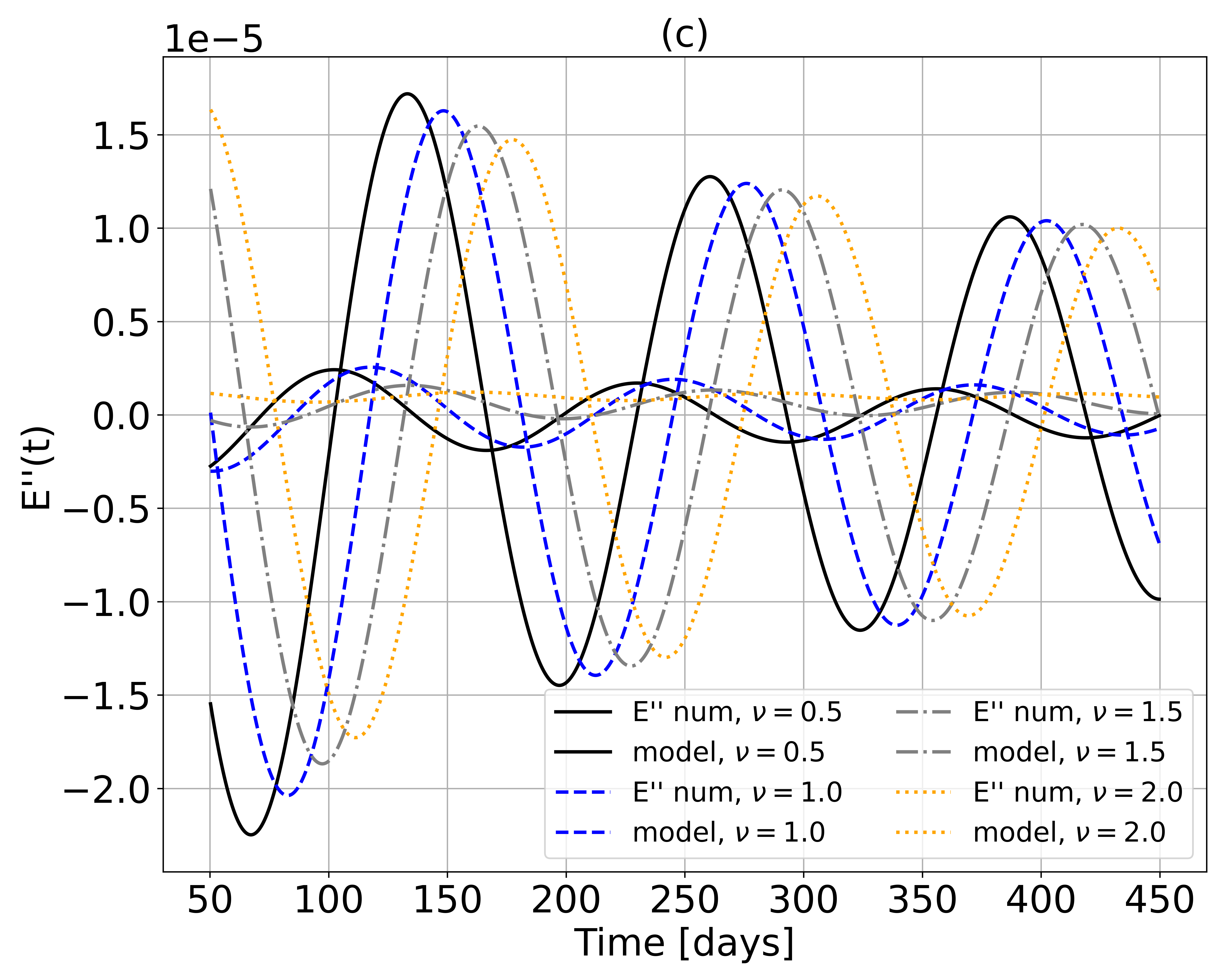}
		\caption{Time evolution of the $^{6}\mathrm{Li}$ enrichment ratio $R(t)$ (a) and the corresponding T breeding ratio $\mathrm{TBR}(t)$ (b) obtained from the linear surrogate model~\eqref{eq:tbr}, together with the target $\mathrm{TBR}_{\text{target}} = 1.0$. Panel (c) shows the second derivative of the T-inventory error $E''(t)$ (solid curves) and the corresponding least-squares second-order fits $E''(t)+a_1E'(t)+a_0E(t)\simeq 0$ (color dashed curves) for several Bessel orders $\nu$. For all cases, the controller maintains small, bounded deviations $0.90 \lesssim \mathrm{TBR}(t) \lesssim 1.10$, and the error dynamics are well described by a low-order LTI model.}
		\label{fig:model_1}
	\end{figure}

	In this simple model, we evolve the control variable $R(t)$ while preserving equation \eqref{eq:n6_n7} and keeping the Li:Pb atomic ratio at $17:83$, as in typical HCLL-like blanket designs, and the total $^6\text{Li} + {}^7\text{Li}$ inventory constant. Under these constraints, the linear surrogate for the T breeding ratio given by equation \eqref{eq:tbr} is used to define the error $E(t) \propto \mathrm{TBR}(t) - \mathrm{TBR}_{\text{target}}$, which is then propagated through the second-order Bessel-type equation derived in Section \ref{sec:pid}. Figure \ref{fig:model_1} shows representative time traces of the $^6\text{Li}$ enrichment $R(t)$ and the corresponding $\mathrm{TBR}(t)$ computed from the linear surrogate model, illustrating that the controller operates in a regime of small, bounded deviations around $\mathrm{TBR} \simeq 1$. For the illustrative examples shown, we adopt $\alpha = 0.08$ and $\beta = 1.0$ in equation \eqref{eq:tbr}, so that the system operates in a regime of small, bounded deviations $0.90 \lesssim \mathrm{TBR}(t) \lesssim 1.10$~  \cite{kuan.1999,zheng.2015,sawan.2006,malang.2009,abdou.2021} around the self-sufficiency point. It is important to stress that for each $\nu$, the corresponding dashed line lies almost on top of the solid one in the chosen time window, which means the closed‑loop error dynamics are well approximated by a second‑order LTI model for all those Bessel orders. Changing $\nu$ shifts the phase and waveform of $E''(t)$, but their amplitude and frequency remain similar, consistent with small, bounded deviations of TBR around the target. Therefore, different “localized Bessel modes” (different $\nu$) still correspond to low‑order, PID‑like dynamics for the T inventory error.
	
	The illustrative values $\alpha = 0.08$ and $\beta = 1.0$ were chosen to fall within the sensitivity ranges reported in DEMO blanket neutronics studies~\cite{sawan.2006,malang.2009,abdou.2021,kuan.1999,zheng.2015}, in which variations of the $^6\text{Li}$ enrichment by a few tens of percent typically induce TBR changes of order $0.05$--$0.1$. In a more detailed setting, $\alpha$ and $\beta$ would be extracted from MCNP/FISPACT calculations for a specific blanket configuration. From an engineering perspective, DEMO-like designs often target TBR values in the range $1.05$--$1.15$, with allowable short-term oscillations of a few percent around the design value~\cite{sawan.2006,malang.2009,abdou.2021,kuan.1999,zheng.2015}. In our surrogate model, the PID--Bessel controller maintains $\mathrm{TBR}(t)$ within approximately $\pm 5\%$ around unity, suggesting that, when retuned for a DEMO-type blanket with a higher design TBR, the same modal structure could be used to keep TBR excursions within comparable engineering tolerances.
	
	Realistic actuation times for modifying the $^6\text{Li}:{}^7\text{Li}$ isotopic enrichment are determined by the processing and circulation loops of the breeding material and are typically much longer than characteristic plasma timescales, spanning from several hours to multiple days~\cite{abdou.2021,kuan.1999}. In the present reduced-order framework, enrichment control is therefore represented as an effectively slow state variable, and the analysis is restricted to the modal structure of the closed-loop error dynamics. A more comprehensive controller synthesis would need to explicitly incorporate these actuation-time limitations and their interactions with other slow subsystems, such as T extraction and overall T inventory management.

	\section{Jet Thermal and PID-Bessel Controller}\label{sec:jet_control}
	
	To illustrate and check the internal consistency of the theoretical framework developed in the preceding sections, we conducted a numerical experiment in Python that integrates a reduced jet/blanket model with the previously described PID--Bessel computational framework. The jet thermal response is modeled using a lumped-parameter temperature field $T(t)$, which obeys a first-order relaxation equation derived by analogy with the coupled mass and heat transport system (4)--(7). In this lumped description, $T(t)$ obeys a first-order relaxation equation of the form
	\begin{eqnarray}
		\frac{\mathrm{d}T}{\mathrm{d}t} = -\frac{1}{\tau_{\mathrm{th}}}\,\bigl(T(t) - T_{\text{eq}}\bigr) + \frac{1}{C_{\mathrm{th}}}\,q_{\text{in}}(t),
	\end{eqnarray}
	where $\tau_{\mathrm{th}}$ is an effective thermal relaxation time, $T_{\text{eq}}$ is the equilibrium temperature in the absence of beam modulation, $C_{\mathrm{th}}$ is a lumped thermal capacitance, and $q_{\text{in}}(t)$ is the Bessel-modulated heat source. This ordinary differential equation is integrated in time using a standard explicit first-order scheme (equivalent to the forward Euler method) with a time step chosen such that $\Delta t \ll \tau_{\mathrm{th}}$, ensuring that the dominant relaxation dynamics are accurately resolved. 
	
	The evolution of $T(t)$ is driven by a Bessel-modulated heat source $q_{\text{in}}(t)$, constructed to emulate the temporally structured energy deposition associated with energetic D beams. The reduced jet/blanket thermohydraulic model, the continuous-time PID law given by equations (8)--(9), and the Bessel mapping of equations (18)--(24) were implemented in a single Python module using standard scientific libraries (NumPy/SciPy)\footnote{The Python scripts used to generate Figs.~3 and~4, including the implementation of the reduced jet/blanket model and the PID--Bessel operator mapping, will be made available as supplementary material (or in a public repository) upon publication, enabling reproducibility and further exploration of alternative gain and parameter choices.}, so that the operator-theoretic mapping can be applied directly to the simulated error trajectories.
	
	The resulting closed-loop error response is subsequently approximated by a second-order model, whose coefficients are then mapped onto effective Bessel parameters using the previously derived relations. This procedure explicitly demonstrates that a physically motivated Li-based system, such as the IFMIF-DONES liquid-Li jet or a fusion blanket surrogate, can exhibit local error dynamics that are accurately characterized by a Bessel-type operator using PID control.
	
\begin{figure}
		\centering
		\includegraphics[width=8.0cm,height=5.0cm]{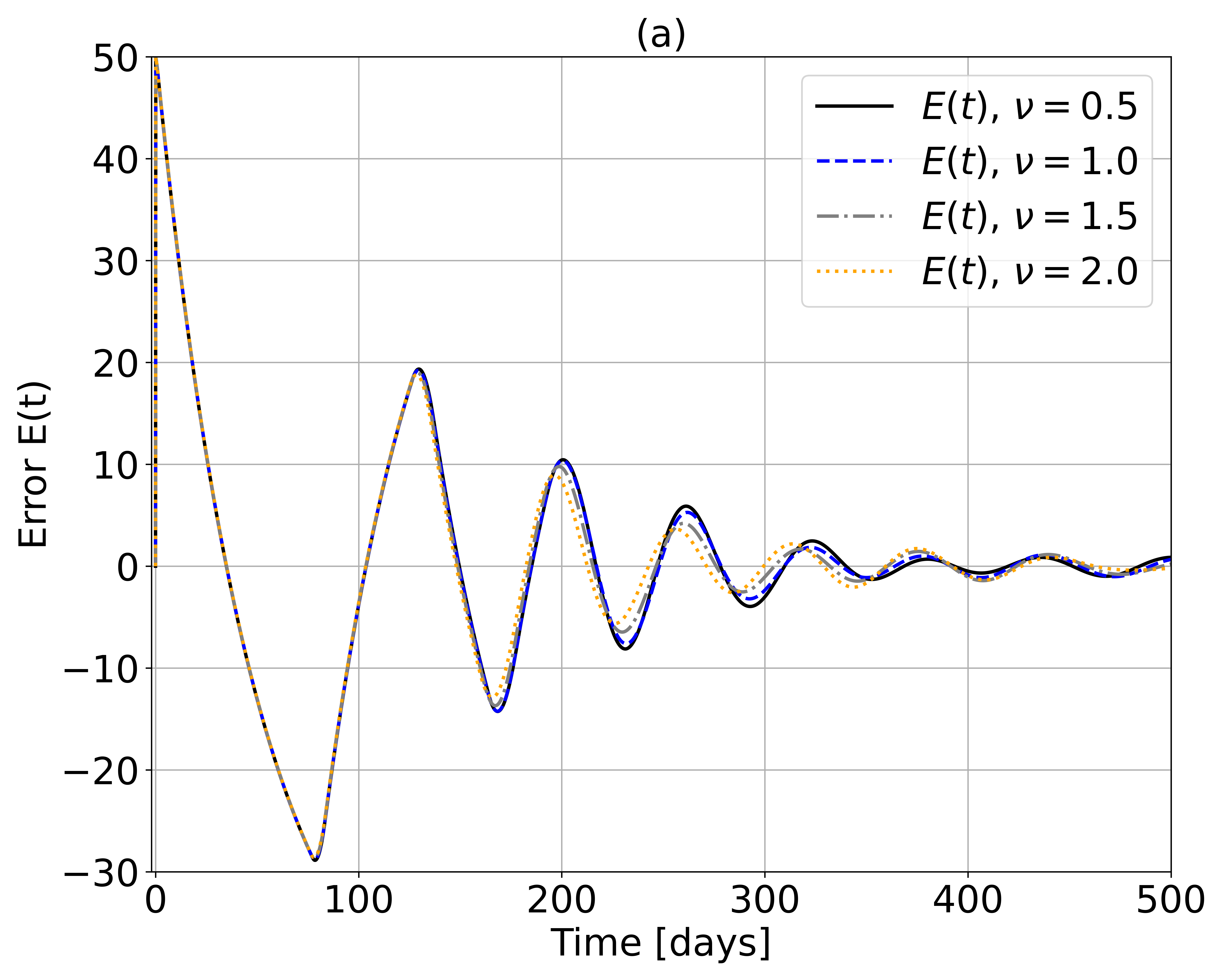}
		\includegraphics[width=8.0cm,height=5.0cm]{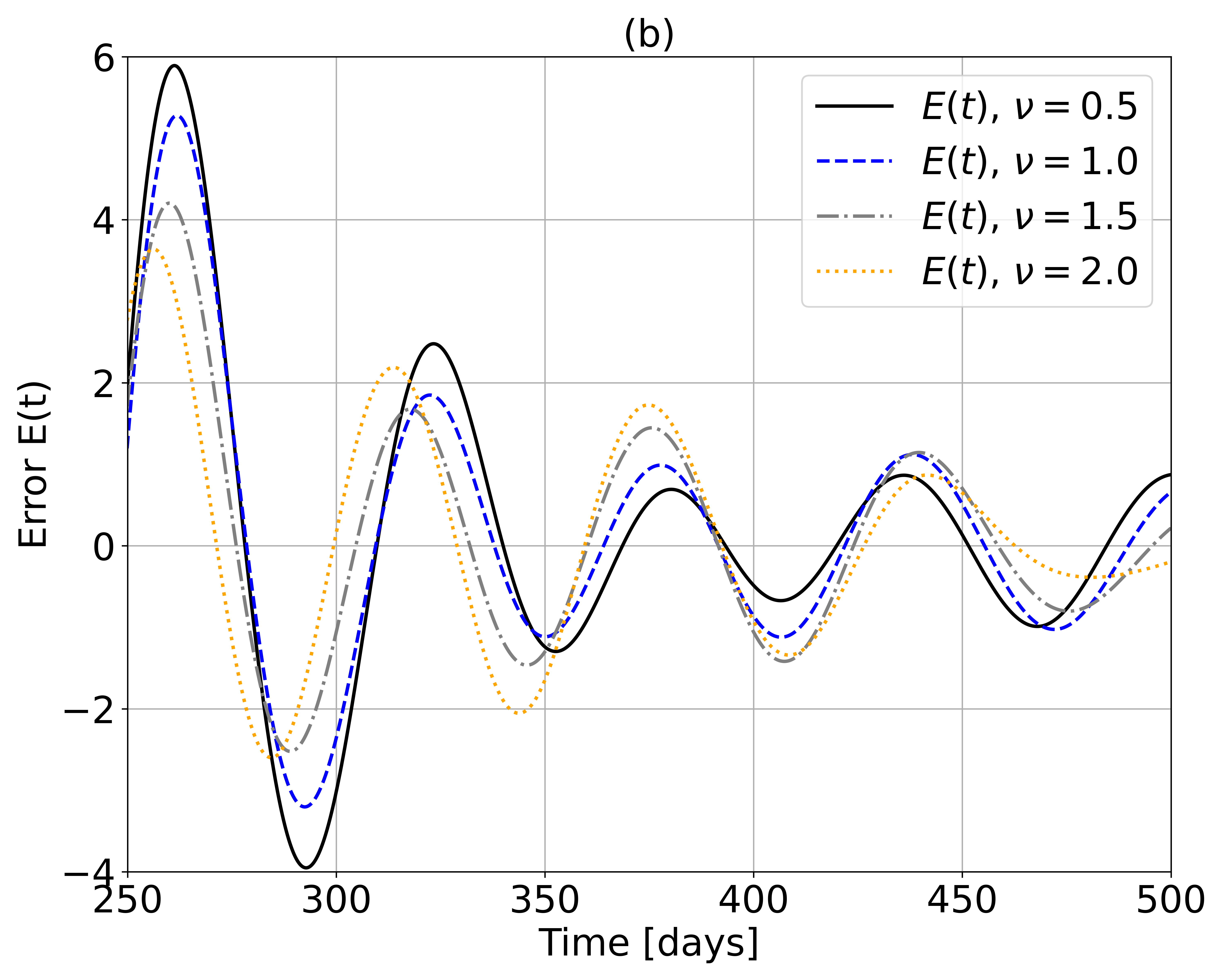}
		\includegraphics[width=8.0cm,height=5.0cm]{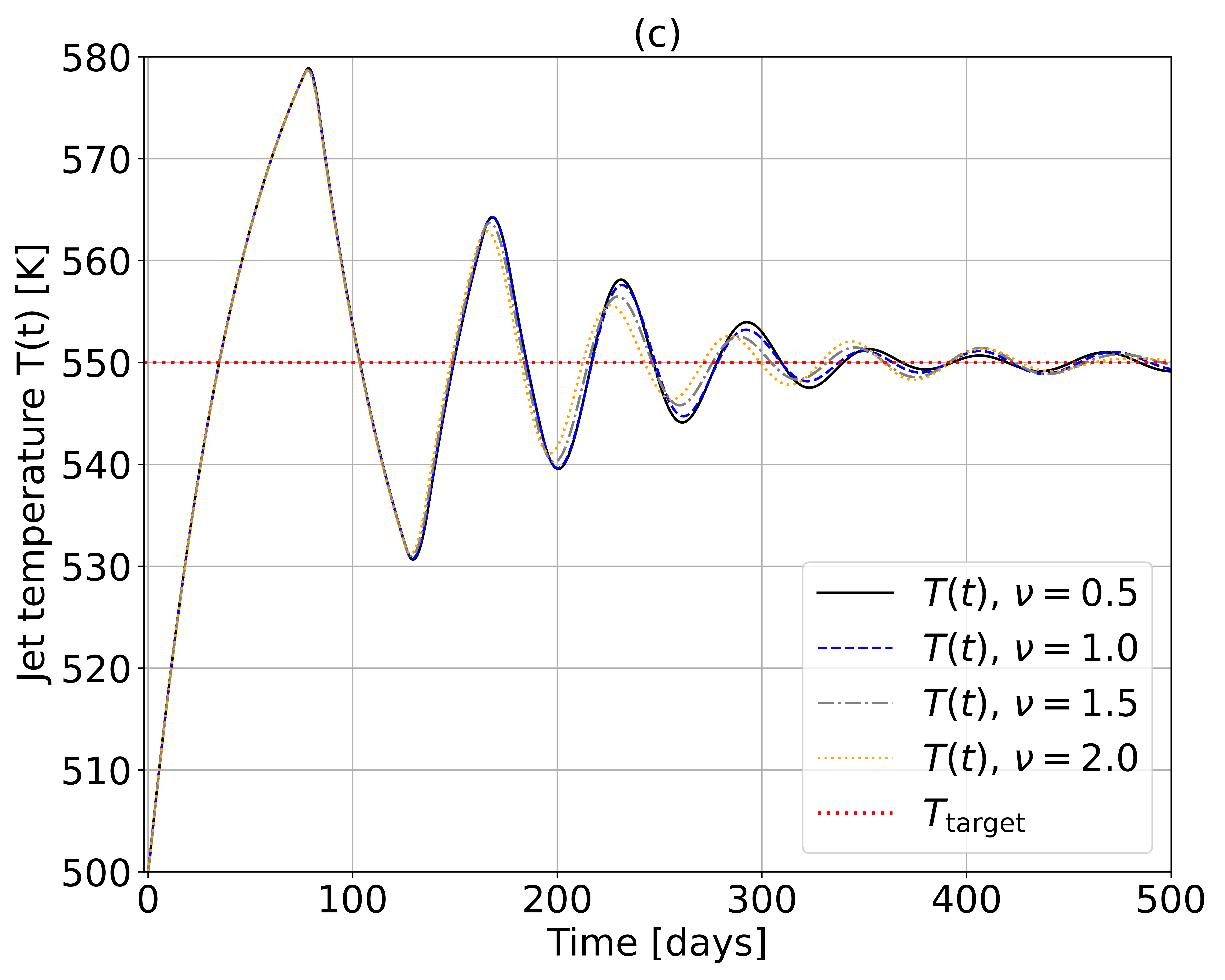} 
		\includegraphics[width=8.0cm,height=5.0cm]{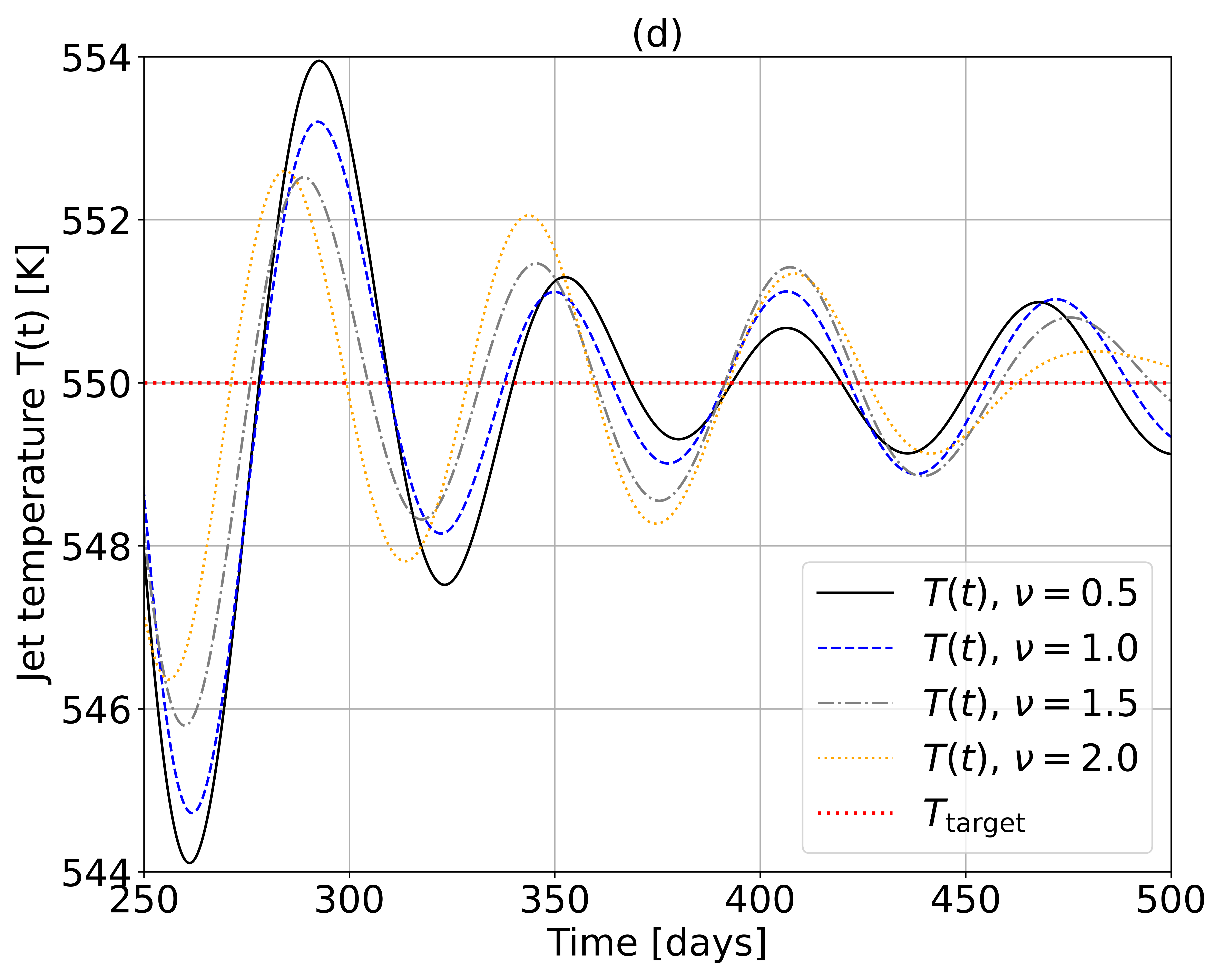} 
		\caption{Panels (a) and (b): numerically computed second derivative of the T‑inventory error $E(t)$ from the reduced jet/blanket–PID model (black solid curves) and corresponding least‑squares fits to the second‑order approximation $E''(t)+a_1E'(t)+a_0E(t)\simeq0$ (color dashed curves) over the approximately linear time window. The close agreement indicates that the closed‑loop error dynamics are well captured by a low‑order model compatible with the localized Bessel‑type operator. Panels (c) and (d) show that, for different $\nu$ (Bessel orders), the closed‑loop jet temperature always relaxes smoothly to the same target value with only mild oscillations.}
		\label{fig:jet}
	\end{figure}

	Figure \ref{fig:jet}, left panel, shows the numerically computed second derivative of the T inventory error $E(t)$ (solid lines) together with the least squares second‑order approximation $E''(t)+a_1E'(t)+a_0E(t)\simeq0$ 
	(dashed lines) for several values of $\nu$. The close overlap between numerical and fitted curves indicates that, over the chosen time window, the closed‑loop error dynamics are well described by a low‑order linear model compatible with the localized Bessel‑type operator. The right panel shows the corresponding jet temperature trajectories $T(t)$ for different Bessel‑modulated heat‑source orders $\nu$. It demonstrates that, despite the differing transient oscillations, the PID controller robustly drives the system toward the same thermal operating point near $T_{\mathrm{targe}}$, confirming that a range of effective Bessel‑type source modulations remains compatible with stable, well‑controlled jet temperatures.

	\section{Discussion and Final Remarks}\label{sec:disc}
	
	Lithium-based components in D-T fusion systems both breed T and remove heat in tokamak blankets and D-Li neutron sources. We combine nuclear data for D-T and Li reactions, simplified thermohydraulic modeling of a liquid Li jet, and an operator-theoretic feedback-control description to reveal the coupling between neutronics, thermal management, and control in these systems.
	
	The jet thermal-expansion model, based on mass conservation and heat conduction, demonstrates that pronounced beam loading induces temperature and velocity perturbations that can be quantitatively characterized by a small set of scalar observables, such as the peak temperature and the maximum transverse velocity. These observables are natural error variables for feedback design and support low-order dynamical models, as in our lumped jet-temperature example driven by a Bessel-modulated heat source.
	
	Writing the continuous-time PID law as a linear operator and comparing it with Bessel differential recurrences exposes a structural analogy. Localizing the Bessel operator around a reference point and introducing a simple second-order error model yields an explicit mapping between PID gains $(a,b,c)$ and effective Bessel parameters $(x_0,\nu)$. Thus, the PID increment can be seen as a localized Bessel-type operator on the T-inventory error. This identification is local but still allows the interpretation of specific PID tunings by selecting effective Bessel modes near an operating point.
	
	In contrast to prior studies that primarily addressed numerical PID tuning for T breeding regulation, this work provides an analytical formulation in which the PID gains are embedded within a Bessel-type operator, thereby elucidating the modal structure of low-order Li-based feedback systems.
	
	Several limitations of this work suggest natural directions for future research. First, the thermohydraulic and TBR models are intentionally simplified and cannot replace full 3D Monte Carlo and Computational Fluid Dynamics (CFD) simulations; the PID–Bessel correspondence should be tested with higher-fidelity models and more realistic operating scenarios. Second, because our construction relies on linearization and local approximation, the mapping is expected to hold only near a chosen operating point; extending the analysis to explicitly nonlinear controllers or gain-scheduled strategies could broaden its validity. Third, we have considered only scalar error variables. In contrast, practical systems involve multiple coupled control loops (e.g., T inventory, outlet temperature, impurity concentration), motivating a multivariable generalization of the Bessel-type operator framework.
	
	Despite these limitations, our results offer an initial step toward a unified treatment of T breeding, thermal management, and feedback control in Li-based fusion technologies. By linking D-T reaction physics, Li-breeding chemistry, jet thermohydraulics, and PID-Bessel operator theory, this work provides a conceptual framework to guide future numerical studies with detailed neutronics and CFD, as well as experiments in facilities such as IFMIF-DONES and next-generation tokamak blankets.
	
	We emphasize that all numerical experiments presented in this study were conducted on a conventional desktop workstation equipped with an Intel\textsuperscript{\textregistered} Core\texttrademark{} i7-class CPU (8 cores, 16 threads) and 16~GB of RAM, running a recent Linux distribution. The reduced jet thermohydraulic simulations (Fig.~2) and the lumped jet/blanket--PID simulations (Figs.~3 and~4) exhibit wall-clock times on the order of a few seconds per realization, without the use of parallelization. The computational tools were implemented in Python and relied on standard scientific libraries (NumPy/SciPy), which are adequate given the low dimensionality and relatively modest computational cost of the reduced-order models considered in this work.
	
	We emphasize that the coupled jet/blanket configuration examined in this study is deliberately idealized and is not intended to represent a realistic power-plant design in which an IFMIF-DONES–type neutron source is directly integrated with a DEMO-class T-breeding blanket operating in a regime close to T self-sufficiency. Rather, it provides a compact analytical framework for exploring low-order, PID-controllable error dynamics in Li-based breeding and heat-removal systems. In future work, a natural direction is to apply the PID–Bessel operator framework to more realistic DEMO-relevant blanket configurations and dedicated materials-testing facilities, using high-fidelity neutronics and CFD simulations as the basis for reduced-order models, while also embedding the present framework into fully three-dimensional jet/blanket geometries with realistic nozzle and duct dimensions. Another extension is to couple additional physics that act on operational timescales, including corrosion of structural materials in contact with liquid Li, T transport and inventory diffusion, magnetohydrodynamic effects, and impurity build-up in breeder and coolant loops, and to generalize the operator framework to multivariable controllers acting simultaneously on T inventory, outlet temperature, and impurity concentrations. These extensions may clarify how the modal structure identified in this work can be systematically translated into robust controller architectures for engineering-scale systems, thereby reinforcing the connection between the present analytical framework and high-fidelity neutronics and CFD investigations.
	
	\section*{Acknowledgments}
	
	The acknowledgments are extended to UFSCar by SASB and SDC for their contributions.
	
	\section*{Disclosure statement}
	
	The authors declare that they have no known competing financial interests or personal relationships that could have appeared to influence the work reported in this paper. No potential conflict of interest was reported by the authors.
	
	\appendix
	\section*{Appendix A: Verification of the Reduced Thermohydraulic Model}\label{sec:appendix}
	
	In this appendix, we present a summary of the quantitative verification tests conducted for the reduced thermohydraulic model introduced in Section \ref{sec:jet}. The purpose of this analysis is to complement the qualitative discussion in the main text by providing explicit error estimates and convergence rates. In doing so, we show that the numerical implementation of equation \eqref{eq:heat} reproduces analytical solutions with controlled accuracy and exhibits the expected spatial and temporal convergence characteristics.
	
	\subsection*{A.1 One-dimensional pure-diffusion test problem}
	
	We first consider a one-dimensional, pure-diffusion test problem with spatially uniform volumetric heat generation and Dirichlet boundary conditions. For this configuration, the semi-discrete formulation of the heat-conduction equation possesses a closed-form analytical steady-state solution, denoted by \(T_{\text{ana}}(z)\). Specifically, we consider a planar slab of thickness \(L\), subject to prescribed (Dirichlet) temperature boundary conditions at the inlet and outlet faces, and exposed to a uniform volumetric heat generation rate \(q_0\) throughout the domain. Under these assumptions, the steady-state analytical temperature profile can be written in the form
	\begin{eqnarray}
		T_{\text{ana}}(z) = T_0 + a z + b z^2,
	\end{eqnarray}
	where the coefficients \(a\) and \(b\) are determined by the boundary conditions and by the balance between diffusion and volumetric heating. This configuration is characteristic of the diffusion-dominated regime relevant to the reduced jet model and serves as a convenient reference solution for verification.
	
	The numerical implementation of equation \eqref{eq:heat} uses second-order central finite differences in space and a first-order explicit scheme in time, with the time step \(\Delta t\) chosen to satisfy the Courant–Friedrichs–Lewy (CFL) stability condition. To quantify the accuracy of this discretization, we compute the numerical steady-state solution \(T_{\text{num}}(z)\) on a sequence of uniform grids and evaluate the relative errors
	\begin{eqnarray}
		\varepsilon_{L_2} = \frac{\|T_{\text{num}} - T_{\text{ana}}\|_{L_2}}{\|T_{\text{ana}}\|_{L_2}},
		\qquad
		\varepsilon_{L_\infty} = \frac{\|T_{\text{num}} - T_{\text{ana}}\|_{L_\infty}}{\|T_{\text{ana}}\|_{L_\infty}},
	\end{eqnarray}
	where \(\|\cdot\|_{L_2}\) and \(\|\cdot\|_{L_\infty}\) denote the standard discrete \(L_2\) and \(L_\infty\) norms over the grid points. Table~\ref{tab:diffusion_errors} presents representative discretization error values corresponding to grid spacings \(\Delta z = 1.0~\mathrm{mm}\), \(0.5~\mathrm{mm}\), and \(0.25~\mathrm{mm}\). For each spatial resolution, the time step \(\Delta t\) is reduced proportionally so as to satisfy the CFL stability condition.
	\begin{table}[h]
		\centering
		\caption{Relative errors for the one-dimensional pure-diffusion verification problem, computed with respect to the analytical steady-state solution $T_{\text{ana}}(z)$.}
		\label{tab:diffusion_errors}
		\begin{tabular}{ccc}
			\hline\hline
			$\Delta z$ (mm) &$\varepsilon_{L_2} = \|T_{\text{num}} - T_{\text{ana}}\|_{L_2} / \|T_{\text{ana}}\|_{L_2}$& $\varepsilon_{L_\infty} = \|T_{\text{num}} - T_{\text{ana}}\|_{L_\infty} / \|T_{\text{ana}}\|_{L_\infty}$ \\
			\hline
			1.00 & $1.10\times 10^{-14}$ & $1.50\times 10^{-14}$ \\
			0.50 & $1.21\times 10^{-14}$ & $2.44\times 10^{-14}$ \\
			0.25 & $2.95\times 10^{-14}$ & $4.51\times 10^{-14}$ \\
			\hline\hline
		\end{tabular}
	\end{table}
	
	In addition to the diffusion test summarized here, we performed systematic grid-refinement analyses for both pure-diffusion and convection–diffusion configurations. For the canonical diffusion case with uniform volumetric heating and Dirichlet boundary conditions, the discrete operator reproduces the analytical steady-state profile with relative errors at the level of machine precision (Table~\ref{tab:diffusion_errors}), so that direct estimation of the convergence order is dominated by roundoff effects. In more general settings with non-uniform sources and boundary conditions, the same finite-difference and time-integration schemes yield relative errors at the percent level and exhibit the expected scaling with mesh size and time step, consistent with second-order spatial accuracy and first-order temporal accuracy. These systematic checks provide additional confidence in the numerical implementation of the reduced thermohydraulic model.
	
	\subsection*{A.2 Convection–diffusion benchmark}
	
	To further test the behavior of the numerical scheme in regimes dominated by advective transport, we consider a canonical convection–diffusion benchmark problem characterized by a prescribed inflow temperature, mixed (Robin-type) thermal boundary conditions at the outflow, and a volumetric heat source confined to a finite interaction region. This configuration is constructed to emulate the beam–matter interaction zone of the IFMIF-DONES liquid-metal jet discussed in Section~\ref{sec:jet}, while preserving a simplified geometry that is well suited to systematic grid-refinement investigations.
	\begin{table}[h]
		\centering
		\caption{Spatial refinement study for the convection–diffusion benchmark: relative $L_2$ error at $t = t_\text{end}$ with respect to a numerically converged reference solution.}
		\label{tab:spatial_refine}
		\begin{tabular}{ccc}
			\hline\hline
			$N_z$ & $\Delta z$ (m) & $\varepsilon_{L_2}$ \\
			\hline
			101  & $5.0\times 10^{-4}$ & $7.26\times 10^{-2}$ \\
			201  & $2.5\times 10^{-4}$ & $3.83\times 10^{-2}$ \\
			401  & $1.25\times 10^{-4}$ & $1.88\times 10^{-2}$ \\
			801  & $6.25\times 10^{-5}$ & $8.25\times 10^{-3}$ \\
			\hline\hline
		\end{tabular}
	\end{table}
	
	For the convection–diffusion benchmark, we performed a sequence of simulations with decreasing $\Delta z$ and $\Delta t$ and compared the resulting temperature fields with a numerically converged reference solution on a highly refined grid. Table~III summarizes the numerical results obtained. In this case, the upwind discretization of the advective term yields an overall spatial accuracy close to first order, as reflected in the observed scaling of the $L_2$ error with mesh size. Nonetheless, the error levels remain at the few-percent level over the parameter ranges of interest, and the reduced thermohydraulic model reproduces the expected linear (diffusion-dominated) and exponential (advection-dominated) temperature profiles, providing a physically reasonable benchmark for the simplified jet configuration.



\begin{thebibliography}{99}
		\bibitem{iaea.2025}International Atomic Energy Agency Bulletin. Fusion Energy. \url{https://www.iaea.org/sites/default/files/fusionenergy.pdf}
		
		\bibitem{ciampichetti.2002}A. Ciampichetti, P. Rocco, and M. Zucchetti. Accidental and Long-Term Safety Assessment of Fission and
		Fusion Power Reactors. Fusion Engineering and Design 63-64, 229-234 (2002).
		
		\bibitem{nevins.1998}W. M. Nevins. A Review of Confinement Requirements for Advanced Fuels. Journal of Fusion Energy 17(1), (1998).
		
		\bibitem{mitei.2024}MIT MITEI. The Role of Fusion Energy in a Decarbonized Electricity System, (2024): \url{https://energy.mit.edu/wp-content/uploads/2024/09/MITEI_FusionReport_091124_final_COMPLETE-REPORT_fordistribution.pdf}
		
		\bibitem{hu.AAPPS.2023}J. Hu {\it et al.}. All Superconducting Tokamak: EAST. AAPPS Bulletin 33, 8 (2023): \url{https://doi.org/10.1007/s43673-023-00080-9}
		
		\bibitem{ding.nature.2024}S. Ding {\it et al.}. A High-Density and High-Confinement Tokamak Plasma Regime for Fusion Energy. Nature 629, 555-560 (2024).
		
		\bibitem{villari.2025}R. Villari {\it et al.}. Overview of Deuterium-Tritium Nuclear Operations at JET. Fusion Engineering and Design 217, 115113 (2025).
		
		\bibitem{Solano.Plasm.Phys.2025}
		E. R. Solano. Fusion Research in a Deuterium–Tritium Tokamak. Fund. Plasma Phys. 15, 100096 (2025).
		
		\bibitem{bosch_hale_1992}H.-S. Bosch and G. M. Hale. Improved Formulas for Fusion Cross-Sections and Thermal Reactivities.  Nucl. Fusion 32(4), 611 (1992).
		
		\bibitem{Souza.arxiv.2019}
		R. S. de Souza \emph{et al.}. Thermonuclear Fusion Rates for Tritium Deuterium Using Bayesian Methods. arXiv:1901.04857 (2019). 
		
		\bibitem{sawan.2006}M. E. Sawan and M. A. Abdou. Physics and Technology Conditions for Attaining Tritium Self-Sufficiency for the DT Fuel Cycle. Fusion Engineering and Design 81, 1131–1144 (2006).
		
		\bibitem{malang.2009}L. A. El-Guebaly and S. Malang. Toward the Ultimate Goal of Tritium Self-Sufficiency: Technical Issues and Requirements Imposed on ARIES Advanced Power Plants. Fusion Engineering and Design 84, 2072–2083 (2009).
		
		\bibitem{abdou.2021}M. Abdou {\it et al.}. Physics and Technology Considerations for the Deuterium–Tritium Fuel Cycle and Conditions for
		Tritium Fuel Self Sufficiency. Nucl. Fusion 61, 013001 (2021).
		
		\bibitem{lee.1972}J. D. Lee. Tritium Breeding and Direct Energy Conversion. Proc. of the  Am.  Chem.  Soc. Symposium  on  The  Role  of   Chemistry  in   the  Development  of  Controlled  Fusion. Boston,  Massachusetts,  (1972): \url{https://inis.iaea.org/records/ehtrj-62m95/preview/4037886.pdf?include_deleted=0} 
		
		\bibitem{manek.2023}P. Mánek {\it et al.}.
		Fast Regression of the Tritium Breeding Ratio in
		Fusion Reactors. Mach. Learn.: Sci. Technol. 4, 015008 (2023).
		
		\bibitem{morgan.2013}L. Morgan and J. Pasley. Tritium Breeding Control Within Liquid Metal Blankets. Fusion Engineering and Design 88, 107–112 (2013).
		
		\bibitem{flament.1992}T. Flament, P. Tortorelli, V. Coen, and H. U. Borgstedt. Compatibility of Materials in Fusion First Wall and Blanket
		Structures Cooled by Liquid Metals. Journal of Nuclear Materials 191-194, 132-138  (1992).
		
		\bibitem{fukada.2010}S. Fukada, Y. Edao, and A. Sagara. Effects of Simultaneous Transfer of Heat and Tritium Through Li–Pb or Flibe Blanket. Fusion Engineering and Design 85, 1314–1319 (2010).
		
		\bibitem{loarte.tech.report.2024}A. Loarte {\it et al.}. Initial Evaluations in Support of the New ITER Baseline and Research Plan. Report no. ITR-24-004, (2024): \url{https://www.iter.org/sites/default/files/media/2024-04/itr-24-004-baseline-ok.pdf}
		
		\bibitem{pitts.nucl.mat.2025}R. A. Pitts {\it et al.}. Plasma-Wall Interaction Impact of the ITER Re-Baseline. Nuclear Materials and Energy 42, 101854 (2025): \url{https://www.sciencedirect.com/science/article/pii/S2352179124002771?via%3Dihub}
		
		\bibitem{HCLL_DEMO}J. Aubert {\it et al.}. Status of the EU DEMO HCLL Breeding Blanket Design Development. Fusion Engineering and Design 136(B), 1428-1432 (2018).
		
		\bibitem{HCPB_DEMO}F. A. Hernandez {\it et al.}. Overview of the HCPB Research Activities in EUROfusion. IEEE Transactions on Plasma Science
		IEEE Transactions on Plasma Science 46(6), 2247-2261 (2018).
		
		\bibitem{WCLL_DEMO}G. Zhou, Y. Lu, and F. A. Hernández. A Water Cooled Lead Ceramic Breeder Blanket for European DEMO. Fusion Engineering and Design 168, 112397 (2021).
		
		\bibitem{morgan.2013_2}L. Morgan and J. Pasley. The Impact of Time Dependant Spectra on Fusion Blanket Burn-up. Fusion Engineering and Design 88, 100-105 (2013).
		
		\bibitem{clark.2025}D.W. S. Clark {\it et al.}. Breeder Blanket and Tritium Fuel Cycle Feasibility of the Infinity Two Fusion Pilot Plant. J. Plasma Phys. 91, E86 (2025). 
		
		\bibitem{kuan.1999}W. Kuan and M. A. Abdou. A New Approach for Assessing the Required Tritium Breeding Ratio and Startup Inventory in
		Future Fusion Reactors. Fusion Technology 35, 309 (1999).
		
		\bibitem{zheng.2015}S. Zheng T. N. Todd. Study of Impacts on Tritium Breeding Ratio of a Fusion DEMO Reactor. Fusion Engineering and Design 98–99, 1915-1918 (2015).
		
		\bibitem{hadi.2026}M. R. Hadi, Md T. Ahmed, M. Rahman, and A. R. Antor. Comparative analysis of tritium breeding ratio of a tokamak reactor using different blanket materials. Annals of Nuclear Energy 236, 112386 (2026).
		
		\bibitem{knaster.2016}J. Knaster {\it et al.}. IFMIF, the European–Japanese Efforts Under the Broader Approach Agreement Towards a Li(d,xn) Neutron Source: Current Status and Future Options. Nuclear Materials and Energy 9, 46-54 (2016).
		
		\bibitem{dovidio.2023}G. D’Ovidio, F. Martín-Fuertes, J. C. Marugan, S. Bermejo, and F. S. Nitti. Lithium Fire Protection Design Approach in IFMIF-DONES Facility. Fusion Engineering and Design 189, 113446 (2023).
		
		\bibitem{hassanein.1996}A. Hassanein. Deuteron Beam Interaction with Lithium Jet in a Neutron Source Test Facility. Journal of Nuclear Materials 233-237, 1547-1551 (1996).
		
		\bibitem{moslang.2008}A. M\"oslang. IFMIF: the Intense Neutron Source to Qualify Materials for Fusion Reactors. C. R. Physique 9, 457–468  (2008). 
		
		\bibitem{Astrom_2002}K. J. \AA str\"om and R. M. Murray. Feedback Systems: An Introduction for Scientists and Engineers, 2nd. edition, (2020). 
		
		\bibitem{Wang.Comm.Phys.2019}
		H.~Wang {\it et al.}. Enhancement of Element Production by Incomplete Fusion Reaction with Weakly Bound Deuteron. Commun. Phys. \textbf{2}, 78 (2019).
		
		\bibitem{wielunska_2016}B. Wielunska, M. Mayer, T. Schwarz-Selinger, U. von Toussaint, and J. Bauer. Cross Section Data for the $D(^3
		\mathrm{He},p)^4\mathrm{He}$ Nuclear Reaction from 0.25 to 6 MeV. Nucl. Instrum. and Meth. in Phys. Research B 371, 41 (2016).
		
		\bibitem{arfken_livro}G. B. Arfken and H. J. Weber. Mathematical Methods for Physicists, 6th edition. Harcourt, (2005).
		
		\bibitem{jiang.2025}Y. Jiang, S. Adulojua, and S. Smolentsev. Design and Analysis of the Open-Surface Slow Li Flow Divertor and Comparison
		to Fast Li Flow Divertor. Fusion Science and Technology 82(1–2), 135–155 (2026).
		
		\bibitem{sizyuk.2022}V. Sizyuk and A. Hassanein. Liquid lithium as divertor material to mitigate severe damage of nearby components during plasma transients. Scientific Rept. 12, 18782 (2022).
		
	\end{thebibliography}
\end{document}